\documentclass[aps,prd,superscriptaddress,showpacs,nofootinbib]{revtex4}
\usepackage{latexsym}
\usepackage{amsmath}
\usepackage{amssymb}

\usepackage{supertabular} 
\usepackage{placeins}
\usepackage{epsfig}
\usepackage{graphicx}
\usepackage{graphics}

\usepackage{xcolor,soul}

\usepackage{url}
\usepackage{hyperref} 
\hypersetup{
   colorlinks=true,
   linkcolor=blue,
   citecolor=blue,
   final=true,
   urlcolor=blue
}

\usepackage{ulem}

\begin{document}
\title{Contact potentials in presence of a regular finite-range interaction using dimensional regularization and the $N/D$ method}
\author{D.R. Entem}
\email[]{entem@usal.es}
\affiliation{Grupo de F\'isica Nuclear and Instituto Universitario de F\'isica 
Fundamental y Matem\'aticas (IUFFyM), Universidad de Salamanca, E-37008 
Salamanca, Spain}

\author{J. Nieves}
\email[]{jmnieves@ific.uv.es}
\affiliation{Instituto de F\'\i sica Corpuscular (CSIC-Universitat de Val\`encia)\\
Parc Cient\'\i fic UV, C/ Catedr\'atico Jos\'e Beltr\'an, 2, 46980 Paterna, Spain}

\author{J.A. Oller}
\email[]{oller@um.es}
\affiliation{Departamento de F\'isica, Universidad de Murcia, E-30071 Murcia, Spain}

\begin{abstract}
We solve the Lippmann–Schwinger equation (LSE) with a kernel that includes a regular finite-range potential and additional contact terms with derivatives. We employ distorted wave theory and dimensional regularization, as  proposed  in Physics Letters B 568 (2003) 109. We analyze the spin singlet nucleon-nucleon $S-$wave as case of study, with the regular one-pion exchange (OPE) potential in this partial wave and up to ${\cal O}(Q^6)$ (six derivatives) contact interactions. We discuss in detail the renormalization of the LSE, and show that the scattering amplitude solution of the LSE fulfills exact elastic unitarity and inherits the left-hand cut of the long-distance OPE amplitude. Furthermore, we proof that the LSE amplitude coincides with that obtained from the exact $N/D$ calculation, with the appropriate number and typology of subtractions to reproduce the effective range parameters taken as input to renormalize the LSE amplitude. The generalization to higher number of derivatives is straightforward.
\end{abstract}

\maketitle

\section{Introduction}
One of the most important aspects on Quantum Field Theory (QFT) is the renormalization of the theory. It has been shown to be
a precise way to compute QFT observables with great success when applied to Quantum Electrodynamics (QED). This theory has the
great advantage that  is perturbative, and calculations as a series on the small coupling
constant $\alpha$ had an unprecedented accuracy~\cite{KinoshitaQED}. However, infinities from high-energy components
of the theory appear, and the renormalization program has to be applied. This scheme on QED works incredibly well.

After QED, with the discovery of the pion, a QFT for strong interaction in terms of nucleons and pions was the obvious
way to go \cite{Yukawa:1935xg}. This was tried by Japanese groups \cite{Taketani:1951,10.1143/ptp/7.1.45,Matsuyama:1958,Hara:1959,Taketani:1960}, and very soon the non-renormalizabilty problem of the theory came into place.
For this reason, the pion theories were abandoned. Later, the quark model was discovered, and the color degree of freedom
of the quarks was introduced. This allowed to build Quantum Chromodynamics (QCD) in perfect analogy to QED as a gauge theory,
but now with a non-abelian $SU(3)$ group of colors. The theory was simple and very important, it was renormalizable.
However, the non-abelian character of the theory that produces interaction between gluons makes it highly non-trivial
and even after more than 50 years working with QCD, we do not have  a completely successful way to treat QCD yet, at
least in all energy regimes.

One of the consequences of the non-abelian character of QCD is that we have a running coupling constant $\alpha_s$ that
becomes small at higher energy scales. For high energy processes this allows to make perturbative calculations as in QED,
but with a slower convergence and much more complicated. However, for low energies, $\alpha_s$ becomes larger and for hadronic scales of the order of 1 GeV, perturbative calculations are not suitable. For this reason, for these systems direct QCD calculations have to be treated numerically on the lattice with important difficulties.

In the last decades Quantum Effective Field Theories (EFTs) had become very popular. The reason is that one does not expect to have theories valid on all energy regimes and even QCD is sometimes seen as an EFT. An EFT provides a quantum-field Lagrangian valid on some energy regime based on the effective degrees of freedom relevant at this scale. Sometimes an EFT is proposed as an effective way to describe the underlying theory. One of the first examples was given by a theory for the interaction of photons proposed by 
Euler and Heisenberg~\cite{Heisenberg:1936nmg}
with QED as underlying theory. An important aspect is that the known exact and approximate symmetries of the underlying theory  are systematically incorporated in the EFT.

However, QFTs have sometimes  a peculiar behavior. If a symmetry is present, the theory can have degenerate vacua, and when
one selects a particular one,  the symmetry is spontaneously broken. 
For massless quarks, QCD is chiral symmetric, but this symmetry is spontaneously broken since degenerate parity-doublets of hadrons are not observed in nature.  EFTs of QCD that include this effect are Chiral Effective Field Theories ($\chi$EFT) and are now
very popular. The consequence of this effect is the appearance of Goldstone (or pseudo-Goldstone) bosons that for
$\chi$EFT are pions and we come back again to theories with pions.

When an EFT is built, very soon terms that are non-renormalizable in the usual way appear and counter-terms to renormalize
the theory are needed. However, using a so called power-counting allows to build a theory that is renormalizable order by order. This has been applied to pion theories with great success.

When one applies these ideas to the nucleon-nucleon ($NN$) system, they do not work. Weinberg realized that the problem was due to an enhancement of the reducible $s-$channel loop-diagram, which breaks the counting rules. He proposed to apply the power-counting program to the potential (two-particle irreducible amplitude), and then use it as a kernel of the Lippmann-Schwinger equation (LSE)~\cite{Weinberg:1990rz,Weinberg:1991um}. However, it turns out that the same counter-terms necessary for the renormalization of the potential do not necessarily renormalize also the LSE. Physicists have been struggling with this problem for more than 30 years~\cite{Kaplan:1996xu,Kaplan:1998tg,Kaplan:1998we,Birse:2005um,Valderrama:2009ei,PavonValderrama:2011fcz,Long:2011qx,Long:2011xw,Long:2012ve,Epelbaum:2006pt,Epelbaum:2009sd,Machleidt:2010kb,Marji:2013uia,Epelbaum:2020maf,Epelbaum:2017byx,Epelbaum:2017tzp,Epelbaum:2018zli,Epelbaum:2019msl,vanKolck:2020llt,Peng:2024aiz,Yang:2021vxa,Baru:2019ndr,Gasparyan:2022isg,Gasparyan:2021edy}. 

In the renormalization program there are two stages. First,  some regularization method to make the divergent integrals of the theory finite should be adopted, and later,  all counter-terms allowed by the symmetry and counting-rules are adjusted to some observables,  renormalizing in this way the theory. In the second step one removes all the dependence on the regulator function. The easiest way to do that is to take a regulator function
that cuts the integrals at some momentum scale $\Lambda$ or some cutoff radius $r_c$ in coordinate space. The last step is taking the
limit $\Lambda \to \infty$ or $r_c\to 0$, respectively.
If one works
in coordinate space with a Scr\"odinger equation this is done using renormalization with boundary conditions. In momentum
space one uses renormalization with a counter-term. Both methods have been shown to be equivalent~\cite{Entem:2007jg}. The main problem is that
only one renormalization condition (singular attractive case) or none (singular repulsive case) can be used in each uncoupled partial wave \cite{Case:1950an,Frank:1971xx,PavonValderrama:2005gu,PavonValderrama:2005wv,PavonValderrama:2005uj}.

Recently the exact $N/D$ method has been developed~\cite{Entem:2016ipb,Oller:2018zts}. It has been shown that the discontinuity of $T$ in the left-hand cut (LHC) can be obtained from an integral equation, which is always finite even for singular interactions.
Then, the analytical properties of the $T$ matrix are exploited by using dispersive relations for the $N$ and $D$ functions, where $D$ has a discontinuity  in the physical cut or right-hand cut (RHC) and $N$ has a cut for imaginary momentum along the LHC (or negative energy), splitting  in this way the two cuts of the $T$ matrix. The RHC of the $T$ matrix is required by unitarity, but the LHC of $T$ matrix is needed as input. 
It has also been shown that the exact $N/D$ method with one subtraction is equivalent to the previous renormalization methods~\cite{Entem:2016ipb,Entem:2021kvs}.
For renormalization with counter-terms, in the  $\Lambda\to\infty$ limit, only the lowest order counter-term survives
(in the singular attractive case), becoming all
the rest irrelevant operators. However it has been shown that the exact $N/D$ method can use more than one subtraction, and even
that singular repulsive interactions can be renormalized, something that has not been shown to be possible with the other methods.

In this work we will analyze the case of a finite range regular interaction in presence of local contact interactions. We will consider more than one counter-term regulating all the ultraviolet (UV) divergencies within  dimensional regularization (DR).  This was done in Refs.~\cite{Kaplan:1996xu, Nieves:2003uu} in the case of the one-pion-exchange (OPE)  potential for the isoscalar spin-singlet $^1S_0$ $NN$ partial $S-$wave with two counter-terms. Since we consider a singlet partial wave, the one pion exchange (OPE) potential is regular (because the constant can be absorbed by the zero order counter-term) and the divergent integrals arise from the iterations of the counter-terms. We will extend the previous work  and show the relation with the exact $N/D$ method with a given number of subtractions. 

 The most important result of this study is that we demonstrate that the LSE amplitude, obtained by solving a two-particle potential which consists of a regular finite-range potential and additional contact terms with an arbitrary number of derivatives, coincides with that of the exact  $N/D$ calculation, with the appropriate number and typology of subtractions to reproduce the effective range parameters taken as input to renormalize the LSE amplitude. This has never been shown before, and it allows better contextualization of the $N/D$ approach with multiple subtractions, at least for regular long-range interactions, providing quasi-analytical expressions for the $T-$matrix where LHC and RHC are self-evident. Furthermore, this achievement connects with two interesting questions.  First, the design of a consistent power-counting able to simultaneously involve the chiral order employed in the long-distance potential and the number of derivatives considered in the short-distance part of the interaction, and second the systematic improvement of data reproduction, without having to introduce/maintain an UV finite cutoff in the theory. These latter issues are beyond the scope of this work, as they require first understanding the case of non-regular long-range interactions, something we are still studying.
 
This work is organized as follows. Section~\ref{sec:dimreg} will be devoted to the solution of the LSE
with  DR techniques, and there, we will provide  results for contact potentials with more than two terms. In Section~\ref{ExactND}, we will present the exact $N/D$ method with a given number of subtractions.  In Section~\ref{results}, we  will compare results from both methods and in Section~\ref{sec:proof},  we will proof the connection between both schemes. We will finish with some conclusions in Section~\ref{conclusions}.  In addition, there is an appendix where we discuss some results with a sharp cutoff, and the relationship between the DR approaches based on the Schr\"odinger equation and on the LSE developed in Refs.~\cite{Kaplan:1996xu} and \cite{Nieves:2003uu}, respectively.
 
\section{LSE with  DR techniques.}
\label{sec:dimreg}

We are going to review and extend the work in Ref.~\cite{Nieves:2003uu} up to ${\cal O}(Q^6)$. The order here refers to the power of momenta considered in the short-distance potential for a given long-distance interaction, which 
is always  the OPE $^1{S_0}$ potential in this work. We will use the same prescriptions as in the original reference. The key idea of the method is to use Distorted Wave Theory\footnote{ It is important to notice that this is different from ``distorted wave Born approximation" in which a part of the interactions is perturbatively treated in a Born series but in distorted waves, that are the exact solution of the other part of the interaction. In  ``distorted wave theory" the solution is exact and includes all the terms of the Born series computed in the  ``distorted wave Born approximation".} (DWT)~\cite{Barford:2002je} to obtain the solution of a long range interaction ($V_\pi$) plus a short range one ($V_s)$ and regulate the divergent integrals using the  DR scheme\footnote{The framework has recently been successfully used to solve strong contact potentials in the presence of the long-distance Coulomb interaction~\cite{Albaladejo:2025kuv}.}. The latter method only keeps UV logarithmically divergences and discards UV power-like divergent scale-less integrals. Note that the finite pion mass regularizes the infrared behavior of the theory. 

Thus, the $^1S_0$ potential is given by
\begin{eqnarray}
	V(p',p) &=& V_\pi(p',p) + V_s(p',p)
	\\
	V_\pi(p',p) &=& -\alpha_\pi \frac{M_N}{2\pi}\frac{1}{p' p} \log\bigg(\frac{p_+^2+m_\pi^2}{p_-^2+m_\pi^2} \bigg),
	\qquad
	\alpha_\pi = \frac{g_A^2m_\pi^2}{16 \pi f_\pi^2}, \quad p_\pm=p\pm p'\label{eq:vope}
	\\
        V_s(p',p) &=& 
	\sum_{s=0}^{3} \sum_{m=0}^s g_{m,s-m} p'^{2m} p^{2(s-m)}=
        g_{00} + g_{01}(p'^2+p^2) 	+ g_{11} p'^2 p^2 + g_{02}(p'^4+p^4)
        \nonumber \\ &&
        + g_{12}(p'^4 p^2+p'^2 p^4)
        + g_{03}(p'^6+p^6) \label{eq:vs}
\end{eqnarray}
where $p$ and $p'$ are the moduli of the $NN$ relative initial and final three momenta in the center of mass frame, $g_{sm}=g_{ms}$ and the $Q^{2n}$ terms are given by $g_{sm}$ with $s+m=n$. We use the parameters of the original  DR calculation of Ref.~\cite{Nieves:2003uu}, $m_\pi= 138$ MeV, $f_\pi= 93$ MeV, $g_A= 1.25$, $M_N=938.919$ MeV and $\hbar c= 197.327$ MeV fm. 

We used a convention such the LSE reads
\begin{eqnarray}
        T(p',p;k^2) &=& V(p',p) + \int_0^\infty dq q^2 \frac{V(p',q) T(q,p;k^2)}{k^2-q^2+i\epsilon} = V + V G_0 T \label{eq:def-lse}
\end{eqnarray}
with $k^2\equiv 2m E=M_NE$, the square of the center of mass momentum and $m=M_N/2$ the reduced mass. The relation with the scattering matrix is
\begin{eqnarray}
        S(k)=e^{2i\delta(k)}= 1 -i\pi k T(k)
\end{eqnarray}
with $T(k)\equiv T(k,k;k^2)$, the on-shell $T-$matrix
and the effective range expansion (ERE) is
\begin{eqnarray}
        k \cot \delta(k) &=& -\frac{2}{\pi} T^{-1}(k) + ik = - \frac{1}{a} + \frac 1 2 r k^2 + v_2 k^4 + \sum_{i=3} v_i k^{2i}
	\label{eq:ere}
\end{eqnarray}
The full off-shell $T-$matrix, $T_\pi$,  for the finite-range interaction $V_\pi$ can be readily obtained  by solving  numerically the appropriate LSE
\begin{eqnarray}
        T_\pi &=& V_\pi + V_\pi G_0 T_\pi
\end{eqnarray}
and it leads to the solution for $T$, which reads~\cite{Nieves:2003uu,Barford:2002je}
\begin{eqnarray}
        T &= T_\pi + (I+T_\pi G_0) T_s (I + G_0 T_\pi)
\end{eqnarray}
where $T_s$ satisfies
\begin{eqnarray}
        T_s &= V_s + V_s G_\pi T_s
\end{eqnarray}
with $G_\pi= G_0 + G_0 T_\pi G_0$. The equation for $T_s$ can be solved algebraically since $V_s$ is a separable potential and
we use the ansatz
\begin{eqnarray}
        T_s(p',p;k^2) &=& 
	\sum_{s=0}^{3} \sum_{m=0}^{3} \alpha_{sm} p'^{2s} p^{2m}
        \nonumber \\ &=&
        \alpha_{00} + \alpha_{01}(p'^2+p^2)
        + \alpha_{11} p'^2 p^2
        + \alpha_{02}(p'^4+p^4)
        + \alpha_{12}(p'^4 p^2+p'^2 p^4)
        + \alpha_{03}(p'^6+p^6)\nonumber\\ &&
        + \alpha_{13}(p'^6 p^2+p'^2 p^6) 
        + \alpha_{22} p'^4 p^4
        + \alpha_{23}(p'^6 p^4+p'^4 p^6)
        + \alpha_{33} p'^6 p^6
\end{eqnarray}
also with $\alpha_{sm}=\alpha_{ms}$. Using the same procedure as in the original work of Ref.~\cite{Nieves:2003uu} we get for the on-shell $T$ matrix
\begin{eqnarray}
        T(k) &=& T_\pi(k) + \frac{\left[1+L_0(k)\right]^2}{\widetilde V_s^{-1}(k)+i\pi k/2 -\bar J_0(k) -J_0^R} \label{eq:tsol}
        \\
        \widetilde V_s(k) &=& h_0 + h_1 k^2 + h_2 k^4 + h_3 k^6
\end{eqnarray}
with the coefficients $h_i$ given in terms of the couplings of the short-range interaction,
\begin{eqnarray}
	h_0 &=& g_{00} -2 g_{01} k_\pi^2 + g_{11} k_\pi^4 + 2 g_{02}k_\pi^2(k_\pi^2+m_\pi^2) 
	-2 g_{03} k_\pi^2(k_\pi^4+13 m_\pi^2 k_\pi^2/3 + m_\pi^4)
	-2 g_{12} k_\pi^4(k_\pi^2+m_\pi^2)
        \nonumber \\
	h_1 &=& 2g_{01}  - 2g_{11} k_\pi^2 - 4 g_{02} k_\pi^2 
	+ 2g_{03} k_\pi^2 (3k_\pi^2+13 m_\pi^2/3)
	+2g_{12} k_\pi^2 (3k_\pi^2+m_\pi^2) 
        \nonumber \\
	h_2 &=& g_{11} + 2 g_{02} - 6k_\pi^2 (g_{03}+g_{12})
        \nonumber \\
	h_3 &=& 2 (g_{03}+g_{12}) \label{eq:hs}
\end{eqnarray}
with $k_\pi^2= 2m m_\pi \alpha_\pi$. Notice that this reproduces the result  with two counter-terms obtained in Ref.~\cite{Nieves:2003uu} for 
$g_{00}=g_0$, $g_{01}=g_1$ and all the other $g_{sm}=0$.
In addition,  we use the functions\footnote{To obtain Eq.~\eqref{eq:tsol} we have used that scaleless integrals are zero in DR, which leads to
\begin{eqnarray}
\int_0^{+\infty} dq\, q^2 V_\pi(q,p) &=& k_\pi^2 \nonumber\\
\int_0^{+\infty} dq\, q^4 V_\pi(q,p) &=& k_\pi^2 (p^2-m_\pi^2) \nonumber \\
\int_0^{+\infty} dq\, q^6 V_\pi(q,p) &=& k_\pi^2\left[ (p^2-m_\pi^2)^2-\frac43 m_\pi^2p^2\right]
\end{eqnarray}
The above integrals are particular cases of 
\begin{equation}
    \int_0^{+\infty} dq\, q^{2n} V_\pi(q,p) = \frac{k_\pi^2}{2n}
\sum_{m=0}^{n-1} {2n \choose 2m+1} p^{2m} (-m_\pi^2)^{n-1-m} \label{eq:vpi-rel}
\end{equation}
for $n\ge 1$, and which allows to obtain the $T-$matrix at any order ${\cal O}(Q^{2n})$ of the expansion of the short-range potential $V_s(p',p)$.
} defined in \cite{Nieves:2003uu}
\begin{eqnarray}
\label{eq:j0}
	J_0(k) &=& \int_0^\infty dq' \int_0^\infty dq \frac{q'^2 q^2 T_\pi(q',q;k^2)}{(k^2-q'^2+i\epsilon)(k^2-q^2+i\epsilon)} = J_0(0) + \bar J_0(k)
	\\
	L_0(k) &=& \int_0^\infty dq\ q^2\frac{ T_\pi(k,q;k^2)}{k^2-q^2+i\epsilon} =\left[l_0(k)-i\pi k/2\right]T_\pi(k),\quad  l_0(k)\in \mathbb{R}\label{eq:l0}
\end{eqnarray}
The loop function  $J_0(k)$, in presence of the long-range OPE interaction, is logarithmically divergent with the UV divergence contained in (see Ref.~\cite{Nieves:2003uu} and  also Appendix A 1)
\begin{eqnarray}
    \int_0^\infty dq \int_0^\infty dq' V_\pi(q',q) &=& -M_N\alpha_\pi \int_0^{+\infty}\frac{dq}{q}\left(\frac{\pi}{2}-\arctan\frac{m_\pi}{q}\right) 
    \end{eqnarray}
If this divergent integral is calculated  in DR one obtains
\begin{eqnarray} 
   -\frac{\pi M_N\alpha_\pi}{4}
        \bigg[
                \frac{1}{3-D} -\gamma_E+ \ln(4\pi) - 2\ln\bigg(\frac{m_\pi}{\mu} \bigg) + 1 +\ldots\bigg]
\end{eqnarray}
where the ellipsis refers to possible finite contributions depending on how the extrapolation of the integral into $D$ dimensions is taken,  and on the additional scale $\mu$ introduced in DR. The bare short-distance coupling constants would cancel this spurious dependence on this scale.

Returning to Eq.~\eqref{eq:j0}, we do a subtraction and thus $\bar J_0(k)$ becomes finite with $J_0^R\equiv J_0(0)$.  Thus, ${\rm Re}\bar J_0(k)$ can be numerically computed and,  in addition, the imaginary part of the loop function\footnote{ Actually, following Subsect. II.E of Ref.~\cite{Albaladejo:2025kuv}, one finds that $L_0(k)$ and $J_0(k)$ are determined by the long-distance OPE scattering wave function at the origin $\Psi_k(\vec 0) = 1 +  L_0(k)= 1+ \left[l_0(k)-i\pi k/2\right]T_\pi(k)$, normalized such that for large values of $r$ it takes the form $\Psi_k(\vec r\,)\sim \left(e^{i\vec{k}\cdot\vec{r}}-\frac{\pi}{2} T_\pi(k)\frac{e^{ikr}}{r}\right)$, while  
\begin{equation}
  - i\frac{\pi k}{2}+J_0(k) = \int_0^\infty dq\ q^2\frac{|\Psi_q(\vec 0)|^2}{k^2-q^2+i\epsilon}=   \int_0^\infty dq\ q^2\ \frac{|T_\pi(q)|^2\left[l_0(q)+{\rm Re}T^{-1}_\pi(q)\right]^2}{k^2-q^2+i\epsilon} \label{eq:j0psi0}
\end{equation}
which is true if $V_\pi$ does not produce bound states. Note that the above expression for $J_0(k)$ involves only the half off-shell $T_\pi(q,q'; q^2)$ matrix for all momenta $q$ needed to compute $l_0(q)$, which replaces the information of full off-shell one   $T_\pi(q',q; k^2)$ for the given  momentum $k$,  as it appears in Eq.~\eqref{eq:j0} above. From Eq.~\eqref{eq:j0psi0}, one easily recovers ${\rm Im}\bar J_0(k)$ given in Eq.~\eqref{eq:imJ0}, while
\begin{equation}
{\rm Re} J_0(k) =-\int_0^\infty dq |\Psi_q(\vec 0)|^2- k^2   \int_0^\infty dq \frac{|\Psi_k(\vec 0)|^2-|\Psi_q(\vec 0)|^2}{k^2-q^2}  
\end{equation}
with the first term logarithmically divergent and the second one, UV finite, gives ${\rm Re}\bar J_0(k)$. }
is given by ~\cite{Nieves:2003uu}, 
\begin{equation}
    {\rm Im}\bar J_0(k) = -\left\{\frac{\pi^2k^2}{4}-l_0(k)\left[l_0(k)+2{\rm Re}T^{-1}_\pi(k)\right]\right\} {\rm Im} T_\pi(k)= \frac{\pi k}{2}+\left[l_0(k)+{\rm Re}T^{-1}_\pi(k)\right]^2{\rm Im} T_\pi(k)\,, \quad k\ge 0  \label{eq:imJ0}
\end{equation}
  As shown in the Appendix~\ref{app:ksw}, the ${\cal O}(Q^2)-$LSE amplitude deduced from Eq.~\eqref{eq:tsol} coincides with the amplitude derived in Ref.~\cite{Kaplan:1996xu} by solving the Schr\"odinger equation in coordinate space.

If the OPE potential is turned off, the usual equation $T^{-1}=V_s^{-1}-G_0$ is found, solved within the  DR scheme. For instance, see Refs.~\cite{Oller:2017alp,Alarcon:2021kpx}. 

Once we fix the finite long-range OPE amplitude, the low energy expansion is governed by the series of $\widetilde V_s^{-1}(k)-J_0^R$, 
\begin{eqnarray}
	\widetilde V_s^{-1}(k) - J_0^R &\sim&
        \delta_0 + \delta_1 k^2 + \delta_2 k^4 + \delta_3 k^6 + \delta_4 k^8+ \ldots \label{eq:deltas}
\end{eqnarray}
The coefficients $\delta_i$ are directly related with the low energy parameters $a$, $r$ and $v_i$ introduced in the ERE of Eq.~\eqref{eq:ere}. Thus, we 
can choose them as renormalization low energy constants (LECs) 
\begin{eqnarray}
        \delta_0 &=& \frac{1}{h_0} - J_0^R
        \nonumber \\
        \delta_1 &=& -\frac{h_1}{h_0^2}
        \nonumber \\
        \delta_2 &=& \frac{h_1^2-h_0h_2}{h_0^3}
        \nonumber \\
        \delta_3 &=& \frac{2 h_0h_1h_2-h_1^3-h_3 h_0^2}{h_0^4} 
        \nonumber \\
        \delta_4 &=& \frac{h_1^4-3 h_0 h_2 h_1^2+h_0^2 h_2^2 + 2 h_0^2 h_1 h_3}{h_0^5}
\end{eqnarray}
This shows the result up to order 6 ($n=3$), however lower orders are contained in this result by simply setting 
$g_{sm}=0$ for $s+m>n$. Notice that for some theories we might have more couplings $g_{sm}$ than renormalization
conditions, and thus families of theories with the same on-shell behavior but different off-shell one an be
generated. This is illustrated at ${\cal O}(Q^6)$ in Eqs.~\eqref{eq:tsol}--\eqref{eq:hs}, where we see that the on-shell amplitude depend on four LECs $h_1,h_2,h_3$ and $h_4$, which are given in terms of the six couplings  $g_{00}, g_{01}, g_{11}, g_{02}, g_{12}$ and $ g_{03}$ that appear in the short-range potential $V_s(p',p)$ at this order (see Eq.~\eqref{eq:vs}).

As shown in Ref.~\cite{Nieves:1999bx}, the $T-$matrix of Eq.~\eqref{eq:tsol} has the appropriate RHC, this is to say it fulfills exact elastic unitarity 
\begin{equation}
 {\rm Im}T^{-1}(k)=  {\rm Im}T_\pi^{-1}(k)= \frac{\pi k}{2}\,,\quad k\ge 0  
\end{equation}
which trivially follows from
\begin{equation}
 T^{-1}(k) =  T^{-1}_\pi(k) - \frac{\widetilde V_s(k)\left[l_0(k)+{\rm Re}T^{-1}_\pi(k)\right]^2}{1 +\widetilde V_s(k)\left({\rm Re}T_\pi(k)\left[l_0(k)+{\rm Re}T^{-1}_\pi(k)\right]^2- J_0^R - {\rm Re}\bar J_0(k)\right)}   \,,\quad k\ge 0  \label{eq:congk}
\end{equation}
that is deduced from Eqs.~(16)-(18) of Ref.~\cite{Nieves:2003uu}.

On the other hand, $T(k)$ inherits the LHC of the long-distance OPE $T_\pi(k)$ amplitude and, as it will be shown in Sect.~\ref{sec:proof},  the second term in the right hand side of Eq.~\eqref{eq:tsol} does not have LHC. Then the discontinuity  at the LHC of the full $T$ and the OPE $T_\pi$ amplitudes is identical 
\begin{equation}
 {\rm Im}T(k)=  {\rm Im}T_\pi(k)\,,\quad k^2\le -\frac{m_\pi^2}{4}  
\end{equation}

As previously mentioned, there is a one to one correspondence between the low energy parameters in the ERE expansion of Eq.\eqref{eq:ere} and the renormalization constants
$\delta_i$. We can rewrite  Eq.~\eqref{eq:congk} as 
\begin{eqnarray}
        T^{-1}(k) &=& T^{-1}_\pi(k) - \frac{\left[l_0(k)+{\rm Re}T^{-1}_\pi(k)\right]^2}{{\rm Re}T^{-1}_\pi(k)+g(k)}
        \nonumber \\
        g(k) &=& \widetilde V_s^{-1}(k) - J_0^R - {\rm Re}\bar J_0(k)+\frac{ {\rm Im} \bar J_0(k)}{\tan \delta_\pi(k)} \label{eq:tsol2}
\end{eqnarray}
with $T^{-1}_\pi(k)=-\pi\left[-ik + k\cot \delta_\pi(k)\right]/2$. From Eq.~\eqref{eq:tsol2} we define
\begin{eqnarray}
        \sum_{i=0} v_i k^{2i} &=& \sum_{i=0} v_i^\pi k^{2i} +\, \frac{\sum_{i=0} a_i k^{2i}}{\widetilde V_s^{-1}(k)-J_0^R+\sum_{i=0} b_i k^{2i}}
\end{eqnarray}
where ($v_0=-1/a, v_1=r/2$)
\begin{eqnarray}
        \sum_{i=0} v_i k^{2i} &=& k \cot \delta(k)
        \nonumber \\
        \sum_{i=0} v_i^\pi k^{2i}  &=& k \cot \delta_\pi(k)
        \nonumber \\
        \sum_{i=0} a_i k^{2i} &=& \frac{2}{\pi} (l_0(k)+{\rm Re}T^{-1}_\pi(k))^2
        \nonumber \\
        \sum_{i=0} b_i k^{2i} &=& {\rm Re}T^{-1}_\pi(k)
        - {\rm Re}\bar J_0(k)+\frac{ {\rm Im} \bar J_0(k)}{\tan \delta_\pi(k)}
	\label{defab}
\end{eqnarray}
with $v_i$ and $v_i^\pi$ the low energy expansion parameters of the full and OPE interactions respectively and the parameters $a_i$ and
$b_i$ given in Table~\ref{tab:aibi}. With this expansion we get the relation
\begin{eqnarray}
        \sum_{i=0} \bar v_i k^{2i} &=& \frac{\sum_{i=0} a_i k^{2i}}{\widetilde V_s^{-1}(k)-J_0^R+\sum_{i=0} b_i k^{2i}}
\end{eqnarray}
with $\bar v_i=v_i-v_i^\pi$. The solution is obtained recursively as
\begin{eqnarray}
        \delta_0 &=& -b_0 + \frac{a_0}{\bar v_0}
        \nonumber \\
        \delta_i &=& -b_i + \frac{1}{\bar v_0} \bigg( a_i
        - \sum_{m=0}^{i-1} (\delta_m +b_m) \bar v_{i-m} \bigg),\qquad i\ge 1
\end{eqnarray}
This method allows to obtain the $\delta_i$ LECs from the ERE parameters $v_i$.

\begin{table}
\begin{center}
\begin{tabular}{lrllrllrll}
\hline
\hline
$i$ & $v_i^\pi$ \hspace*{0.0cm} & & & $a_i$ \hspace*{0.0cm} & & & $b_i$ \hspace*{0.0cm} & & \\
\hline
0 &  0.113860 & $10^{}$ & fm$^{-1}$ & $ 0.222872\,$ & $10^{  6}$ & MeV$^{2}$ & $-0.991969\,$ & $10^{  3}$ & MeV         \\
1 &  0.618830 & $10^{}$ & fm & $ 0.438752\,$ & $10^{  2}$ &            & $-0.697427\,$ & $10^{ -1}$ & MeV$^{-1}$  \\
2 & $-$0.836529 & $10^{}$ & fm$^{3}$ & $ 0.434318\,$ & $10^{ -3}$ & MeV$^{-2}$    & $ 0.307092\,$ & $10^{ -5}$ & MeV$^{-3}$  \\
3 &  0.336524 & $10^{2}$ & fm$^{5}$ & $ 0.119093\,$ & $10^{ -7}$ & MeV$^{-4}$    & $-0.329470\,$ & $10^{ -9}$ & MeV$^{-5}$  \\
4 & $-$0.172440 & $10^{3}$ & fm$^{7}$ & $-0.289344\,$ & $10^{-11}$ & MeV$^{-6}$    & $ 0.440900\,$ & $10^{-13}$ & MeV$^{-7}$  \\
5 &  0.995614 & $10^{3}$ & fm$^{9}$ & $ 0.516084\,$ & $10^{-15}$ & MeV$^{-8}$    & $-0.659949\,$ & $10^{-17}$ & MeV$^{-9}$  \\
6 & $-$0.617247 & $10^{4}$ & fm$^{11}$ & $-0.897871\,$ & $10^{-19}$ & MeV$^{-10}$ & $ 0.105708\,$ & $10^{-20}$ & MeV$^{-11}$ \\
7 &  0.401154 & $10^{5}$ & fm$^{13}$ & $ 0.157580\,$ & $10^{-22}$ & MeV$^{-12}$ & $-0.177167\,$ & $10^{-24}$ & MeV$^{-13}$ \\
8 & $-$0.269598 & $10^{6}$ & fm$^{15}$ & $-0.280558\,$ & $10^{-26}$ & MeV$^{-14}$ & $ 0.306710\,$ & $10^{-28}$ & MeV$^{-15}$ \\
9 &  0.185777 & $10^{7}$ & fm$^{17}$ & $ 0.506675\,$ & $10^{-30}$ & MeV$^{-16}$ & $-0.544040\,$ & $10^{-32}$ & MeV$^{-17}$ \\
\hline
\hline
\end{tabular}
\caption{\label{tab:aibi} Expansion coefficients for different quantities, defined in Eq.(\ref{defab}), and involving the half off-shell OPE amplitude. }
\end{center}
\end{table}

\section{The exact $N/D$ method.}
\label{ExactND}

The $N/D$ method was introduced by Chew and Mandelstam in Ref.~\cite{Chew:1960iv}. The basic idea is to separate the cuts of the
scattering amplitude in the following way
\begin{eqnarray}
	T_{ND}(A) &=& \frac{N(A)}{D(A)}
\end{eqnarray}
with $A=k^2$ such that $D(A)$ and $N(A)$ have a RHC and a LHC in the complex $A$ plane, respectively. Note that the normalization of the $T-$matrix in this section is different to that used in Sec.~\ref{sec:dimreg} and both are related by $T_{ND}(A)=-2\pi^2 T(A)/M_N$, with $T(A)$ the on-shell $T-$matrix with complex energy $A/M_N$. We also notice that here we use the usual prescription for the $N/D$ method that differs from the one in Ref.~\cite{Entem:2021kvs}. The RHC is the unitarity cut due to intermediate $NN$ states and the LHC is generated by particle exchanges.  One then uses dispersion relations with integrals along the cuts to obtain the amplitudes in the entire complex plane.

The $T_{ND}(A)$ amplitude is real between the LHC and the RHC, and thus $N(A)$ and $D(A)$ are real in the RHC and LHC, respectively. The discontinuity on the cuts only involve the imaginary parts of the functions. In the RHC the unitarity condition implies
\begin{eqnarray}
	{\rm Im}[D(A)] &=& - N(A) \rho(A), \qquad A>0
\end{eqnarray}
with $\rho(A)=M_N\sqrt A/4\pi$, with the cut of the multi-valued square root function taken from 0 to $+\infty$.  The $\rho$ phase space factor and the $S-$matrix are related by  $S(A)=1+2i\rho(A) T_{ND}(A)$. On the LHC, since $D(A)$ is real, we have
\begin{eqnarray}
	{\rm Im}[N(A)] &=& D(A) {\rm Im}[T_{ND}(A)],\qquad A < -\frac{m_\pi^2}{4}  
\end{eqnarray}
The dispersion relations only involve the imaginary part of the functions $N$ and $D$ on the cuts, and therefore 
${\rm Im}[T_{ND}(A)]$ on the LHC is the only input needed, which encodes all dynamical information. 

The general form of the $N/D$ equations was given in Ref.~\cite{Guo:2013rpa} with different number of subtractions in $N$ and $D$. We need to do at least one subtraction in $N$ or $D$, and we decide to do it in $D$ fixing $D(0)=1$ at the $NN$ threshold. Performing more subtractions allows to fix also derivatives of $D$ and/or $N$ at threshold and $N(0)$, which is equivalent to fix the low energy parameters in the ERE of Eq.~(\ref{eq:ere}). Thus by fixing the scattering length $a$, one gets the integral equations
\begin{eqnarray}
	D(A) &=& 1+i \sqrt{A} a
	+i\frac{M_N}{4\pi^2} \int_{-\infty}^L d\omega_L \frac{D(\omega_L)\Delta(\omega_L)}{\omega_L} \frac{A}{\sqrt{A}+\sqrt{\omega_L}}
\nonumber \\
	N(A) &=& -\frac{4\pi a}{M_N} + \frac{A}{\pi} \int_{-\infty}^L d\omega_L \frac{D(\omega_L)\Delta(\omega_L)}{(\omega_L-A)\omega_L}
	\label{ND11}
\end{eqnarray}
with the LHC limit $L=-m_\pi^2/4$  and $2i\Delta(A)=2i {\rm Im}[T_{ND}(A)]$  the discontinuity of the $T$ matrix on the LHC. One first solves the integral equation for $D(A)$ on the LHC and with it, $N(A)$
and $D(A)$ can be evaluated on the whole complex plane. An arbitrary number of subtractions can be performed, allowing in principle to fix the lowest terms  of the ERE to some given values. We will denote as
$N/D_{nd}$, the $N/D$ calculation with $n$ subtractions in $N$ and $d$ subtractions in $D$, fixing the $n+d-1$ lowest effective range parameters $a,r,v_2,\ldots,v_{n+d-2}$. When the $N/D$ solution exists, it is unique once the number of subtractions has been fixed regardless of whether they were done in $N$ or $D$, but the convergence of the integral equations might depend on the number of subtractions performed in the $N$ and $D$ functions.

Then to use the $N/D$ method one needs $\Delta(A)$ on the LHC and  it has been traditionally evaluated perturbatively. The exact $N/D$ method
was introduced in Ref.~\cite{Entem:2016ipb} and developed in Ref.~\cite{Oller:2018zts} giving an integral equation to obtain $\Delta(A)$ from the discontinuity of the potential. The general equation for orbital angular momentum $l$ reads 
\begin{equation}
        \Delta \hat T_{ND}(\nu,\bar k) = \Delta \hat V(\nu,\bar k) + \theta(\bar k-m_\pi)
        \theta(\bar k-2m_\pi-\nu) \frac{M_N}{2\pi^2} \int_{\nu+m_\pi}^{\bar k-m_\pi} d\nu_1 \,\frac{\nu_1^2}{\bar k^2-\nu_1^2} S(\nu_1)
        \Delta \hat V(\nu,\nu_1) \Delta \hat T_{ND}(\nu_1,\bar k)
        \label{IntEq}
\end{equation}
with $k=\sqrt A= i\bar k$, and
\begin{eqnarray}
        S(\nu_1) &=& \frac{1}{(\nu_1+i\epsilon)^{2l+2}} + \frac{1}{(\nu_1-i\epsilon)^{2l+2}}
        \\
        2\Delta V(\nu_1,\nu_2) &=& 
	\frac{2\Delta \hat V(\nu_1,\nu_2)}{\nu_1^{l+1} \nu_2^{l+1}}=\lim_{\epsilon\to 0} \lim_{\delta\to 0}
        \big[
        {\rm Im} V(i\nu_1+\epsilon-\delta,i\nu_2+\epsilon)
        -{\rm Im} V(i\nu_1+\epsilon+\delta,i\nu_2+\epsilon)
        \big]
        \\
        2\Delta T_{ND}(\nu,\bar k) &=& 
	\frac{2\Delta \hat T_{ND}(\nu,\bar k)}{\nu^{l+1} \bar k^{l+1}}= \lim_{\epsilon\to 0} \lim_{\delta\to 0}
        \big[
        {\rm Im} T_{ND}(i\nu+\epsilon-\delta,i\bar k+\epsilon)
        -{\rm Im} T_{ND}(i\nu+\epsilon+\delta,i\bar k+\epsilon)
        \big]
        \label{250522.1}
\end{eqnarray}
where the discontinuity on the LHC is
\begin{eqnarray}
        \Delta (A) &=& -\frac{\Delta \hat T_{ND}(-\bar k,\bar k)}{\bar k^{2l+2}}
\end{eqnarray}
Notice that the integral equation Eq.(\ref{IntEq}) is always finite and no cutoff scale is introduced.

 We will apply the $N/D$ method to the $^1S_0$ $NN$ long-range finite OPE interaction $V_\pi(p',p)$ of Eq.~\eqref{eq:vope}. The required LHC discontinuity $\Delta(A)$  was discussed and obtained in  Ref.~\cite{Entem:2016ipb}. The $N/D_{01}$ calculation, which does not need any input from the ERE expansion, provides $T_\pi(k)$. In addition, $N/D_{nd}$ calculations, with $n+d\ge 2$, can be used to find solutions with some ERE parameters different than those found with the finite OPE potential. 

\section{Renormalization of the  DR amplitudes }
\label{sec:renorm}

In the discussion of the results of the  DR approach, we will distinguish two different renormalization schemes, depending on whether $J^R_0$ is finite or infinite. Both of them lead to renormalized amplitudes that differ only in higher-order terms than those considered in the short-range contact potential.

The $J^R_0\to \infty$ scheme ($DR_{\infty}$ scheme) can be straightforwardly applied to 
any order ${\cal O}(Q^{2n})$ and it provides a systematic and improvable reproduction of the ERE. Within this scheme, all $\delta_i$ observables, introduced in Eq.~\eqref{eq:deltas},  beyond the considered order $n$ are zero, and therefore, the renormalization procedure does not induce any correlation between higher-order $\delta_{i>n}$,  as would occur as a result of keeping the UV subtraction constant $J_0^R$ finite. 

The second renormalization scheme, where $J^R_0$ is kept finite, is more phenomenologically successful since, at given order ${\cal O}(Q^{2n})$, the subtraction constant can be adjusted to describe the next ERE-order $(k^{2n+2})$ to that considered in the expansion of the short-range contact potential. Moreover, there exist some renormalization-group equations which provide the dependence of the bare parameters ($g_{sm}$, with  $s+m\le n$) of $V_s$ on $J^R_0$ to guaranty that  the $\delta_{0,1,\cdots,n}$ LECs are independent of the  phenomenological finite value of $J^R_0$ employed to reproduce $\delta_{n+1}$. This scheme would be also systematically improvable and was followed in the original  DR paper of Ref.~\cite{Nieves:2003uu}. Actually, its philosophy is similar to that behind the renormalization procedures employed   in the pioneering works on the $f_0(500)$, $f_0(980)$ and $a_0(980)$ resonances \cite{Oller:1997ti}, on the $\Lambda(1405)$  \cite{Oset:1997it} and its double pole nature  \cite{Oller:2000fj}, 
or on the heavy-quark spin symmetry partners of the $X(3872)$ \cite{Guo:2013sya}\footnote{This procedure has been adopted in many other works about these resonances and on the study of many other dynamically generated states. It is also illustrating to pay attention to the $\pi\pi$ scattering analysis carried out in  Ref.~\cite{Nieves:1999bx} (see also \cite{Nieves:1998hp}), where the Bethe-Salpeter equation (BSE) in $S-$wave is solved using the leading order (LO) chiral perturbative potential for all isospin channels. There, it is shown the existing relation between subtraction constants, needed to render UV finite the amplitudes, and the next-to-leading (NLO) chiral  physical (renormalization independent) LECs $\bar \ell_{1,2,3,4}$ introduced by Gasser and Leutwyler in Ref.~\cite{Gasser:1983yg}. In particular, the so-called {\it on-shell} scheme discussed in Sect. 4 of Ref.~\cite{Nieves:1999bx} amounts to consider the loop-function UV logarithmic divergences in each $\pi\pi$ isospin channel, while all power-like UV divergencies are discarded, replacing the effects of the latter by the appropriate renormalization of  the bare two-particle irreducible potential used as kernel of the BSE. The loop-function logarithmic divergencies are similar to $J_0^R$ here. In the LO BSE calculation of Ref.~\cite{Nieves:1999bx}, they were obtained from the phenomenological numerical values of the Gasser-Leutwyler NLO $\bar \ell_{1,2,3,4}$ LECS, which is in good approximation equivalent to fit the experimental scattering length and effective range in each isospin sector. A difference approach was taken in Ref.~\cite{Oller:1997ti}, where they were numerically estimated  by cutting off the momentum integration above certain $q_{\rm max}$ of around 1 GeV.}.  However, at a given order ${\cal O}(Q^{2n\ge 4})$, this renormalization scheme cannot  always be adopted depending of the specific values taken by the  $\delta_{0,1,\cdots,n,n+1}$ LECs. This is because it may even be necessary to set  $J_0^R$ to a complex value, which would spoil unitarity. The problem appears since beyond $v_2$ (or equivalently $\delta_2$) precision in the ERE, one needs to solve a nonlinear equation to determine the UV subtraction constant $J_0^R$  from the $\delta_{0,1,\cdots,n,n+1}$ observables. Indeed, one finds that $\delta_{n+1}$ is given by a polynomial of rank $n$ in the variable $(\delta_0+J_0^R)^{-1}$ with coefficients determined by the lower-order  $\delta_{0,1,\cdots,n}$ quantities. One might find more than one real solution for $J_0^R$, giving rise to different theories that will provide different ERE beyond the $p^{2n+2}$ term. 
If all solutions for $J_0^R$ are complex, one should go the next order ${\cal O}(Q^{2n+2})$ and see if  the set of $(n+3)$ observables $\delta_{0,1,\cdots,n,n+1, n+2}$ can be reproduced with the enlarged bare potential and a real subtraction constant $J_0^R$. Otherwise, one can fix $J_0^R$ to certain real value  and tune the bare $V_s$ couplings to reproduce the $\delta_{0,1,\cdots,n,n+1}$ LECs. These latter $(n+2)$ observables  can be also exactly described
within the $J^R_0\to \infty$ scheme at  ${\cal O}(Q^{2n+2})$ order.

To finish the discussion of this section, we note that traditional dimensional regularization scheme (see for instance Refs.~\cite{Collins:1984xc,Kaiser:2011cg,Phillips:1997xu,Schafer:2005kg}) would correspond to the   DR procedure, but keeping $J_0^R$ finite and fixed by the election of some reasonable renormalization scale or virtual momentum cutoff~\cite{Kaplan:1996xu} (see also the discussion in Appendix~\ref{app:cut}).

\section{Results.}
\label{results}

We are going to compare the results of the  DR calculation of Sec.~\ref{sec:dimreg} with the result of the
exact $N/D$ method of Sec.~\ref{ExactND} order by order in the expansion of the contact potential $V_s(p',p)$ [Eq.~\eqref{eq:vs}].

\subsection{Order 0, $n=0$}
\label{order0}

At first order the short-distance potential is only the constant $g_{00}$ 
($g_{sm}=0$ for $s+m>0$).

Though in principle, there appear two renormalization constants, $g_{00}$ and $J_0^R$, however there is only one
renormalization condition
\begin{eqnarray}
	\widetilde V_s^{-1}(k) - J_0^R\Big|_{{\cal O}(Q^0)} &=& \frac{1}{g_{00}} - J_0^R = \delta_0 \quad \Rightarrow  \quad 
	g_{00} = \frac{1}{J_0^R + \delta_0}
\end{eqnarray}
and the result only depends on $\delta_0$, which can be determined from the scattering length. The  value $a=-23.7588$~fm leads  to $\delta_0/ m=-0.0811$.  On the other hand, one finds a renormalization equation
\begin{equation}
 \frac{dg_{00}}{dJ_0^R}=- g_{00}^2  
\end{equation}
which constraints how  the bare coupling should approach zero ($g_{00}\to 0$) in the $J_0^R\to \infty$ limit.
\begin{figure}
\begin{center}
\includegraphics[width=7.5cm]{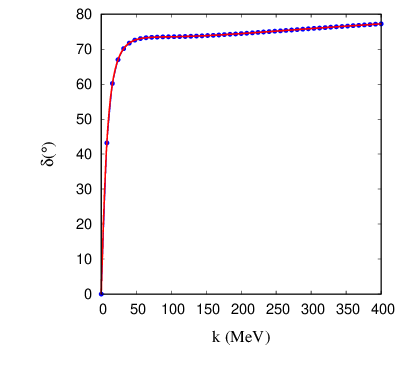}
\caption{\label{forder0} Comparison of the $^1S_0$ phase-shifts obtained at order ${\cal O}(Q^0)-$DR (blue line with dots)
and the non-perturbative $N/D_{11}$ (red line) methods, fixing in both cases the scattering length to the same common value, as a function of the $NN$ center of mass momentum $k$.  One can identify this solution with those obtained in Refs.~\cite{PavonValderrama:2005wv} and \cite{Nogga2005}   using renormalization with boundary conditions and with one counter-term, respectively.}
\end{center}
\end{figure}

The predicted phase-shifts are depicted in Fig.~\ref{forder0}. The blue line with dots shows the result of the
DR calculation while the red line stands for the $N/D$ method result fixing
the scattering length $a$ to the previous value. We can see a perfect agreement between both calculations.
As we know from Refs.~\cite{Entem:2016ipb,Oller:2018zts,Entem:2021kvs} that the $N/D$ method is equivalent to renormalization with boundary conditions and with cutoff renormalization with one counter-term, all the calculations agree. The low energy
parameters obtained in both methods agree and are given up to $v_6$ by\footnote{We use the analytical continuation of $T(k)$ to complex values of $k^2$ and  Cauchy's theorem to compute $d^n(k\cot \delta)/d(k^2)^n\big|_{k^2=0}$.}
\begin{eqnarray}
        a &=& -23.7588 \,{\rm fm}
        \nonumber \\
        r &=&  1.34787 \,{\rm fm}
        \nonumber \\
        v_2 &=& -2.02330 \,{\rm fm^3}
        \nonumber \\
        v_3 &=& 9.25287 \,{\rm fm^5}
        \nonumber \\
        v_4 &=& -50.4842 \,{\rm fm^7}
        \nonumber \\
        v_5 &=& 303.006 \,{\rm fm^9}
        \nonumber \\
        v_6 &=& -1930.40 \,{\rm fm^{11}} \label{eq:EREorden2}
\end{eqnarray}
So at order 0 we find the  DR is equivalent to $N/D_{11}$ fixing the scattering length.

\subsection{Order 2, $n=1$}
\label{order1}

At order ${\cal O}(Q^2)$, a new contact term ($g_{01}$) is added to the short-range potential. This is the case studied in the original  DR calculation of Ref.~\cite{Nieves:2003uu}. Now we have in principle three renormalization constants, namely, $g_{00}$, $g_{01}$ and $J_0^R$. As anticipated, we will consider two different renormalization schemes depending on whether $J_0^R$ is kept finite (DR), and becomes a physical LEC, or is sent to infinity (DR$_\infty$).

\subsubsection{${\cal O} (Q^2)$  DR scheme}
\label{sec:DRorden2}
\begin{figure}
\begin{center}
\includegraphics[width=7.5cm]{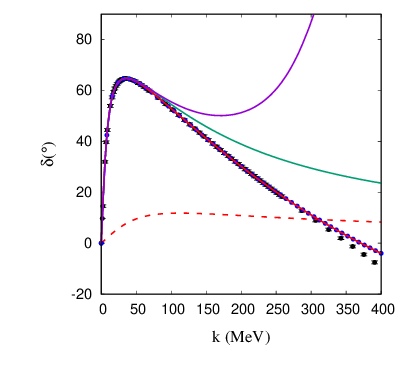}
\includegraphics[width=7.5cm]{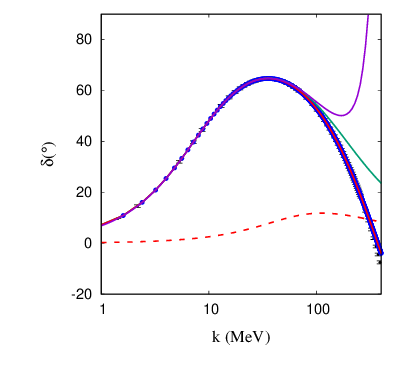}
\caption{\label{fig:forder1} Comparison of the phase-shifts obtained in the ${\cal O}(Q^2)-$DR (blue line with dots)
and the non-perturbative $N/D_{22}$ (red line) methods fixing $a$, $r$ and $v_2$. The green line stands for 
the result of the $N/D_{12}$ solution fixing $a$ and $r$ and the magenta curve shows the prediction from the effective range approximation including terms up to $v_2$.  The black dots with error bars are the experimental phase-shifts from \cite{Stoks:1993tb} used to fit the order 2 ($n=1$) solution. The right panel is the same but using a log-scale for $k$.  Furthermore, the red dashed curves represent the phase-shifts obtained from the long-range OPE potential $V_\pi $.}
\end{center}
\end{figure}

In  Ref.~\cite{Nieves:2003uu}, the three LECs were fitted to the $^1S_0$ $NN$ Nijmegen phase shits~\cite{Stoks:1993tb} with the values
\begin{eqnarray}
	g_{00}\, m &=& -0.2762
	\nonumber \\
	g_{01}\, m^3 &=& 0.3470
	\nonumber \\
	J_0^R/m &=& -3.2070 \label{eq:num-valPLB}
\end{eqnarray}
The ERE parameters for this calculation are
\begin{eqnarray}
        a &=& -23.7588 \,{\rm fm}
        \nonumber \\
        r &=&  2.67286 \,{\rm fm}
        \nonumber \\
        v_2 &=& -0.571396 \,{\rm fm^3}
        \nonumber \\
        v_3 &=& 5.00024 \,{\rm fm^5}
        \nonumber \\
        v_4 &=& -29.2870 \,{\rm fm^7}
        \nonumber \\
        v_5 &=& 185.601 \,{\rm fm^9}
        \nonumber \\
        v_6 &=& -1224.71 \,{\rm fm^{11}}
	\label{eq:lowJuan}
\end{eqnarray}
Since in this scheme there are three independent parameters, we compare to the $N/D_{22}$ solution fixing $a$, $r$ and $v_2$. We find a perfect agreement between the low energy parameters obtained within both non-perturbative methods. The predicted phase-shifts are displayed in Fig.~\ref{fig:forder1}, where we see again a perfect agreement.  As before, the $N/D_{22}$ is shown as a red line and the  DR calculation
as a blue line with dots. In addition, the green and magenta curves stand for the results of the $N/D_{12}$ solution fixing only $a$ and $r$ and of the ERE going up to $v_2$, respectively.

Now we focus on the renormalization details of the ${\cal O}(Q^2)-$DR scheme. Fixing $\delta_0$ and $\delta_1$, these renormalization conditions lead to the couplings.
\begin{eqnarray}
	g_{00}(J_0^R) &=& \frac{1}{\delta_0+J_0^R} - \frac{\delta_1 k_\pi^2}{(\delta_0+J_0^R)^2},  \qquad g_{01}(J_0^R) =
    - \frac{\delta_1}{2(\delta_0+J_0^R)^2} \label{eq:renororden2}
\end{eqnarray}
and all these theories fixes $a$ and $r$. Indeed, there exists a whole family of solutions with bare parameters $g_{00}(J_0^R)$ and $g_{01}(J_0^R)$, functions of the UV subtraction constant $J_0^R$, which provide the same numerical values for the $\delta_0$ and $\delta_1$ observables independently of $J_0^R$. On the other hand, the ${\cal O}(p^4)$ term in the ERE ($v_2$ or equivalently $\delta_2$) at this ${\cal O} (Q^2)$ order can be now used to fix $J_0^R$,
\begin{eqnarray}
	\delta_2 &=& \frac{\delta_1^2}{\delta_0+J_0^R} \label{eq:delta2orden2}
\end{eqnarray}
Finally setting  
\begin{equation}
   J_0^R=j_0^{(Q^2)}= -\delta_0 + \delta_1^2/\delta_2 \label{eq:j0orden2}
\end{equation}
and its numerical value from $\delta_0, \delta_1$ and $\delta_2$ obtained for the ERE parameters $a,r$ and $v_2$, we obtain that for this particular value $j_0^{(Q^2)}$ of the UV subtraction constant 
\begin{eqnarray}
	g_{00}[j_0^{(Q^2)}] &=& \frac{\delta_2(\delta_1 -k_\pi^2 \delta_2)}{\delta_1^3}, \qquad g_{01}[j_0^{(Q^2)}] = -\frac{\delta_2^2}{2\delta_1^3} \label{eq:para-app}\\
	\widetilde V_s^{-1}(k)-J_0^R\Big|_{{\cal O}(Q^2)} &=& \delta_0 + \frac{\delta_1^2 k^2}{\delta_1-\delta_2 k^2} 
	\sim	\delta_0 + \delta_1 k^2 + \delta_2 k^4 + \sum_{m=3}^\infty \frac{\delta_2^{m-1}}{\delta_1^{m-2}} k^{2m}
	\label{eq:serie1}
\end{eqnarray}
and thus once we fix $\delta_0$, $\delta_1$ and $\delta_2$ (or equivalently $a$, $r$ and $v_2$), the theory 
is completely fixed at this order.

From the LECs of Eq.~\eqref{eq:num-valPLB} used in Ref.~\cite{Nieves:2003uu} we get 
\begin{eqnarray}
	\delta_0 &=& -0.08111\, m
	\nonumber \\
	\delta_1  &=& -7.50329\, m^{-1}
	\nonumber \\
	\delta_2  &=& -17.1221\, m^{-3}
	\label{eq:delta13org}
\end{eqnarray}
with these values, the  $a$, $r$ and $v_2$ ERE parameters will be given by Eq.(\ref{eq:lowJuan}) and all higher order LECs 
$\delta_m= \delta_2^{m-1}/\delta_1^{m-2}$ with $m\ge 3$ will be fixed, as deduced from Eq.~\eqref{eq:serie1}. This  is  in principle a spurious correlation, which at this order of the momentum expansion of the contact potential $V_s$, is as valid as setting all $\delta_{m\ge 3}$ LECs to zero.  Some further insights are discussed in the Appendix.
\subsubsection{${\cal O} (Q^2)$ DR$_{\infty}$ scheme}
\label{sec:drinftyorden1}
Now we consider the  limit $J_0^R\to\infty$, which amounts to set $\delta_2=0$ (see Eqs.~\eqref{eq:delta2orden2} and \eqref{eq:j0orden2}), and then
\begin{equation}
	\lim_{J_0^R \to \infty} \widetilde V_s^{-1}(k) - J_0^R = \delta_0 + \delta_1 k^2
\end{equation}
and the observables $\delta_0$ and $\delta_1$ become independent of the regulator $J^R_0$, when it is taking to infinity, thanks to the dependence of the bare couplings $g_{00}$ and $g_{01}$ on $J^R_0$ given in Eq.~\eqref{eq:renororden2}. In this limit we find
\begin{eqnarray}
      g_{00}(J_0^R) &=& \frac{1}{J_0^R}- \frac{\delta_0 + k_\pi^2 \delta_1}{[J_0^R]^2}+ {\cal O}\big(\frac{1}{[J_0^R]^3}\big)+\cdots
	\nonumber \\
 g_{01}(J_0^R) &=& -\frac{\delta_1}{2[J_0^R]^2} +{\cal O}\big(\frac{1}{[J_0^R]^3}\big)+\cdots \label{eq:gJ0Rorden1}
\end{eqnarray}
These are group of renormalization equations which guarantee that $d\delta_0/dJ_0^R$ and $d\delta_1/dJ_0^R$ become of order ${\cal O}(1/[J_0^R]^2)$ and therefore vanish in the $J_0^R\to\infty$ limit.  The explicit inclusion of  the  ${\cal O}\big(1/[J_0^R]^3\big)$ terms in Eq.~\eqref{eq:gJ0Rorden1} would allow to eliminate the ${\cal O}(1/[J_0^R]^3)$ contributions to  $d\delta_0/dJ_0^R$ and $d\delta_1/dJ_0^R$, and so on. In this way it is possible to arbitrarily reduce the residual dependence of the calculated values for the $\delta_0$ and $\delta_1$ observables on a large, but finite, value of $J_0^R$ and to define the $J_0^R \to \infty$ limit of the theory.

\begin{figure}
\begin{center}
\includegraphics[width=7.5cm]{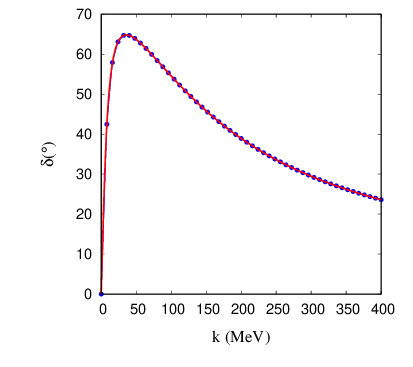}
\caption{\label{ND12} Phase-shifts for the ${\cal O}(Q^2)$ DR$_{\infty}$ scheme (blue line with dots)  
($n=1$) compared with the results from the  $N/D_{12}$ method (red line).} 
\end{center}
\end{figure}

In this scheme, we are also setting $\delta_i=0$ for $i\ge 2$, removing the correlation of the higher terms in Eq.~\eqref{eq:serie1}. Thus, in this renormalization scenario,  we fix $a$ and $r$ and the rest of higher ERE parameters
are also fixed and determined from  $a$ and $r$ and the full off-shell long-range $T_\pi$ matrix. The result of the low energy parameters in this limit are
\begin{eqnarray}
        a &=& -23.7588 \,{\rm fm}
        \nonumber \\
        r &=&  2.67286 \,{\rm fm}
        \nonumber \\
        v_2 &=& -0.838491 \,{\rm fm^3}
        \nonumber \\
        v_3 &=& 4.57626 \,{\rm fm^5}
        \nonumber \\
        v_4 &=& -27.7418 \,{\rm fm^7}
        \nonumber \\
        v_5 &=& 177.297 \,{\rm fm^9}
        \nonumber \\
        v_6 &=& -1176.81 \,{\rm fm^{11}}
\end{eqnarray}
which perfectly agree with the $N/D_{12}$ solution fixing $a$ and $r$. The comparison of the phase-shifts
is shown in Fig.~\ref{ND12}.

Thus at ${\cal O}(Q^2)$, we find that keeping $J_0^R$ finite, the  DR scheme is equivalent to the $N/D_{22}$ fixing $a$, $r$ and $v_2$, while the $J_0^R \to \infty$ limit of the theory is equivalent to $N/D_{12}$ fixing only $a$ and $r$.

\subsection{Order 4, $n=2$}
\label{sec:order2}
At this order,  there appear in the short range potential four LECs $g_{00}$, $g_{01}$, $g_{11}$ and $g_{02}$. However, $g_{02}$ can be eliminated by redefining $g_{00}\to g_{00}+2g_{02}k_\pi^2m_\pi^2$ and $g_{11}\to g_{11}+2g_{02}$, cf. Eq.~\eqref{eq:hs}, as expected since a polynomial of rank 4 in even powers of $k$ has only three independent parameters

\subsubsection{${\cal O} (Q^4)$  DR scheme}

If we fix $\delta_0$, $\delta_1$ and $\delta_2$ the renormalization conditions give the couplings
\begin{eqnarray}
	g_{00}(J_0^R)  &=& -2k_\pi^2 m_\pi^2 g_{02}(J_0^R) +
	\frac{1}{\delta_0+J_0^R} 
	- \frac{(\delta_1+k_\pi^2 \delta_2) k_\pi^2}{(\delta_0+J_0^R)^2} 
	+ \frac{\delta_1^2 k_\pi^4}{(\delta_0+J_0^R)^3} 
	\nonumber \\
	g_{01}(J_0^R) &=& - \frac{\delta_1+2k_\pi^2\delta_2}{2(\delta_0+J_0^R)^2} 
	+ \frac{\delta_1^2 k_\pi^2}{(\delta_0+J_0^R)^3} 
	\nonumber \\
	g_{11}(J_0^R) &=& -2g_{02}(J_0^R)  - \frac{\delta_2}{(\delta_0+J_0^R)^2} +\frac{\delta_1^2}{(\delta_0+J_0^R)^3}
\end{eqnarray}
Here again, there exists a whole family of solutions with bare parameters $g_{00}^{\rm eff}(J_0^R)=g_{00}(J_0^R)+2k_\pi^2 m_\pi^2 g_{02}(J_0^R)$, $g_{01}(J_0^R)$ and $g_{11}^{\rm eff}(J_0^R)=g_{11}(J_0^R)+2 g_{02}(J_0^R)$ functions of the UV subtraction constant $J_0^R$, which provide the same numerical values for the $\delta_0$, $\delta_1$ and $\delta_2$ observables independently of $J_0^R$
 
In addition, at this order $\delta_3$ may be used to determine $J_0^R$ since  
\begin{eqnarray}
	\delta_3 &=& \frac{2\delta_1\delta_2}{\delta_0+J_0^R} 
	- \frac{\delta_1^3}{(\delta_0+J_0^R)^2} \label{eq:delta3Orden2}
\end{eqnarray}
Hence to fix $\delta_3$, we have a second order equation with  solutions
\begin{eqnarray}
	 j_0^{(Q^4_{\pm})}&=& -\delta_0 + \frac{\delta_1}{\delta_3}\bigg(\delta_2 \pm \sqrt{\delta_2^2-\delta_1\delta_3}\bigg)
	\label{eq:J0R2}
\end{eqnarray}
The first thing to notice is that, as mentioned above, we have a one parametric ($g_{02}$) family of theories that are
equal on-shell, however they are different off-shell. The reason is that we only have $h_0$, $h_1$ and
$h_2$ to fix the couplings, so only three equations.

The second thing to notice is that for real $J_0^R$ we only have solutions fixing $\delta_3$ for
\begin{eqnarray}
	 \delta_2^2-\delta_1\delta_3 \ge 0 \label{eq:segundo-orden}
\end{eqnarray}
and we have two different solutions if the inequality in Eq.~\eqref{eq:segundo-orden} is fulfilled. This illustrates the discussion in Sec.~\ref{sec:renorm}.  It is worth remembering at this point that a complex value of $J_0^R$ is not acceptable, as it would not satisfy unitarity.

\begin{figure}
\begin{center}
\includegraphics[width=7.5cm]{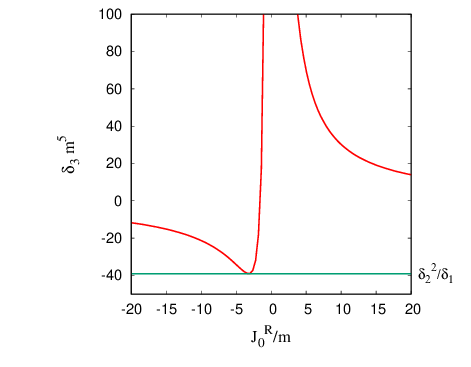}
\caption{\label{fig:delta3} Red line: Dependence of $\delta_3$ on $J_0^R$ as deduced from the ${\cal O}(Q^4)$  Eq.~\eqref{eq:delta3Orden2}, and using  the numerical values of $\delta_0$, $\delta_1$ and $\delta_2$ given in the ${\cal O}(Q^2)$ Eq.~\eqref{eq:delta13org}. The green line stands for the minimum value $\delta_3\big|_{\rm min}=\delta_2^2/\delta_1=-39.0717\,m^{-5}$, which is reached for $J_0^R=(\delta_1^2-\delta_0\delta_2)/\delta_2=-3.2070\, m$ [Eq.~\eqref{eq:num-valPLB}].  }
\end{center}
\end{figure}

Note that in the case of considering the ${\cal O}(Q^2)$ approach ($n=1$), then $\delta_2^2=\delta_1 \delta_3$, as inferred from Eq.~(\ref{eq:serie1}). Actually using the  ${\cal O}(Q^2)$ numerical values given in Eq.~\eqref{eq:delta13org}, we find $\delta_3=\delta_2^2/\delta_1=-39.0717\,m^{-5}$, which is the minimum\footnote{This trivially follows from the dependence of $\delta_3$ on $J_0^R$ exhibited in Eq.~\eqref{eq:delta3Orden2}. That observable has a extreme at $J_0^R=(\delta_1^2-\delta_0\delta_2)/\delta_2$ and for this $J^R_0$, one finds $\delta_3=\delta_2^2/\delta_1$. In addition, $d^2\delta_3/d(J_0^R)^2= -2\delta_2^4/\delta_1^5$ for the above $J_0^R$, which turns out to be positive (minimum) for the numerical values of Eq.~\eqref{eq:delta13org}.} value possible for $\delta_3$ with $J_0^R\in  \mathbb{R}$, as can be seen in Fig.~\ref{fig:delta3}.
For $\delta_1$ and $\delta_2$ given in Eq.~\eqref{eq:delta13org}, numerical values of $\delta_3$ less than $\delta_2^2/\delta_1=-39.0717\, m^{-5}$ are not accessible at this order requiring a real  $J_0^R$. As pointed out in Sec.~\ref{sec:renorm}, any phenomenological $\delta_3$ can be reproduced going to  the next order ${\cal O} (Q^6)$. 

\begin{figure}
\begin{center}
\includegraphics[width=7.5cm]{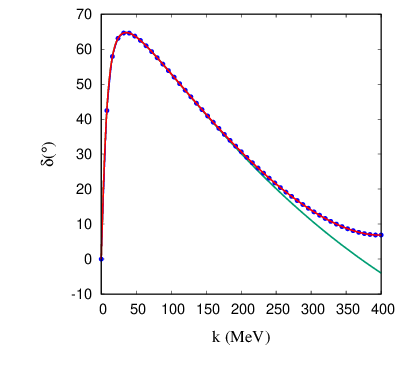}
\includegraphics[width=7.5cm]{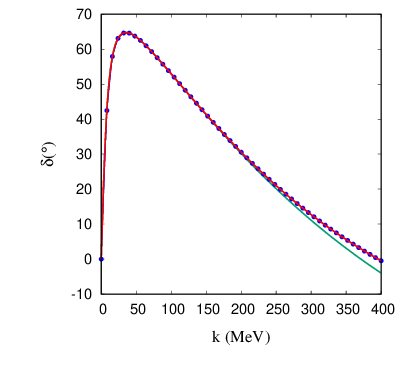}
\caption{\label{fig:deltao2} Phase-shifts obtained from the $N/D_{33}$ (red lines) and  ${\cal O}(Q^4)$ ($n=2$)  DR (blue lines with dots) methods, fixing the ERE parameters up to $v_4$. Scenarios 1 [Eq.~\eqref{eq:sol1}]  and 2 [Eq.~\eqref{eq:sol2}] are shown in the left  and right panels of the figure, respectively. The green curve in both panels stands for the original  DR calculation of Ref.~\cite{Nieves:2003uu} calculated at order  ${\cal O}(Q^2)$ ($n=1$), and as detailed in the text, it has a different value for $v_4$ than the two ${\cal O}(Q^4)$ scenarios depicted in the plots.}
\end{center}
\end{figure}

To better discuss the equivalence with the $N/D$ method at ${\cal O} (Q^4)$, we increase $\delta_3(<0)$ and multiply its modulus by 0.9, which can be obtained with a real value of the subtraction constant $J_0^R$. Thus, we take
\begin{equation}
  \delta_3=  -35.1645\,m^{-5}\label{eq:delta3}
\end{equation}
We get two solutions with the
couplings for solution 1
\begin{eqnarray}
	\left(g_{00} + 2 g_{02} k_\pi^2  m_\pi^2 \right)  m &=& -0.35332
	\nonumber \\
	g_{01} m^3 &=& 0.56625
	\nonumber \\
	\left(g_{11}+ 2g_{02}\right)  m^5 &=&  -0.86761
	\nonumber \\
	J_0^R/m &=& j_0^{(Q^4_+)}= -2.41702
\end{eqnarray}
and for solution 2
\begin{eqnarray}
	\left(g_{00} + 2 g_{02} k_\pi^2  m_\pi^2 \right)  m  &=&  -0.19452
	\nonumber \\
	g_{01} m^3 &=& 0.17166
	\nonumber \\
	\left(g_{11}+ 2g_{02}\right)  m^5 &=&  0.23415
	\nonumber \\
	J_0^R/m &=&  j_0^{(Q^4_-)}=-4.72767
\end{eqnarray}
The two solutions are different although both of them provide the same numerical values of $\delta_0,\delta_1,\delta_2$ and $\delta_3$. In fact we obtain
\begin{widetext}
    \begin{eqnarray}
	\widetilde V_s^{-1}(k)-J_0^R\Big|_{{\cal O}(Q^4)}  &\sim& \delta_0 + \delta_1 k^2 + \delta_2 k^4 + \delta_3 k^6 
	- \frac{\delta_2^3-2\delta_1\delta_2\delta_3\mp (\delta_2^2-\delta_1\delta_3)^{3/2}}{\delta_1^2} k^8
	+\ldots \label{eq:seriesorden2}
\end{eqnarray}
\end{widetext}
where the $\mp$ sign above corresponds to $j_0^{(Q^4_{\pm})}$ in Eq.~(\ref{eq:J0R2}). This means that fixing up to $\delta_3$ (or equivalently up to $v_3$) does not completely determine the
theory and we need to fix up to the term $v_4$ in the ERE. We so perform and $N/D_{33}$ calculation fixing the low energy
parameters up to $v_4$ given by
\begin{eqnarray}
        a &=& -23.7588 \,{\rm fm}
        \nonumber \\
        r &=&  2.67286 \,{\rm fm}
        \nonumber \\
        v_2 &=& -0.57140 \,{\rm fm^3}
        \nonumber \\
        v_3 &=& 4.98949 \,{\rm fm^5}
        \nonumber \\
        v_4 &=& -29.3099 \,{\rm fm^7}
         \label{eq:sol1}
\end{eqnarray}
for solution 1  and
\begin{eqnarray}
        a &=& -23.7588 \,{\rm fm}
        \nonumber \\
        r &=&  2.67286 \,{\rm fm}
        \nonumber \\
        v_2 &=& -0.57140 \,{\rm fm^3}
        \nonumber \\
        v_3 &=& 4.98949 \,{\rm fm^5}
        \nonumber \\
        v_4 &=& -29.3072 \,{\rm fm^7}
        \label{eq:sol2}
\end{eqnarray}
for solution 2. The $N/D_{33}$ and  DR phase-shifts for these two scenarios are shown in Fig.~\ref{fig:deltao2}, where we see an excellent  agreement. To compare to the previous results obtained at order 2 (green curves), we should notice that the modulus of $\delta_3$ used here is 10\% smaller than that induced by  the  correlation in the ${\cal O}(Q^2)$ calculation. This change produces a $v_3(>0)$ in Eqs.~\eqref{eq:sol1} and \eqref{eq:sol2} almost a factor of two smaller than that given in Eq.~\eqref{eq:EREorden2} for the ${\cal O}(Q^2)$ theory. As a consequence, we get at high energies for both ${\cal O}(Q^4)$ solutions a larger attraction than in the  ${\cal O}(Q^2)$ case.  

\subsubsection{${\cal O} (Q^4)$ DR$_{\infty}$ scheme}
\label{sec:drinftyorden2}
In the $J_0^R \to \infty$ limit we get
\begin{eqnarray}
	\lim_{J_0^R\to\infty} 
	\widetilde V_s^{-1}(k)-J_0^R &=& \delta_0 + \delta_1 k^2 + \delta_2 k^4 
	\nonumber \\
	g_{00}(J_0^R) + 2 g_{02}(J_0^R) k_\pi^2 m_\pi^2 &=&  \frac{1}{J_0^R}- \frac{\delta_0 + k_\pi^2 \delta_1 + k_\pi^4 \delta_2}{[J_0^R]^2}+ {\cal O}\big(\frac{1}{[J_0^R]^3}\big)+\cdots
	\nonumber \\
	g_{01} (J_0^R) &=& -\frac{\delta_1+2k_\pi^2\delta_2}{2[J_0^R]^2} + {\cal O}\big(\frac{1}{[J_0^R]^3}\big)+\cdots
	\nonumber \\
	g_{11}(J_0^R) + 2g_{02} (J_0^R) &=&-\frac{\delta_2}{[J_0^R]^2} + {\cal O}\big(\frac{1}{[J_0^R]^3}\big)+\cdots \label{eq:gJ0Rorden2}
\end{eqnarray}
The discussion here is analogous to that in Sec.~\ref{sec:drinftyorden1}. The relations of Eq.~\eqref{eq:gJ0Rorden2} are group of renormalization equations at order ${\cal O}\big(1/[J_0^R]^2\big)$ and guarantee  that $d\delta_0/dJ_0^R$, $d\delta_1/dJ_0^R$ and $d\delta_2/dJ_0^R$ become of order ${\cal O}(1/[J_0^R]^2)$, and hence these vanish in the $J_0^R\to\infty$ limit. The residual dependence of the computed $\delta_{0,1,2}$ observables on $J_0^R$ can be systematically removed to any required precision.

We are now fixing $\delta_i=0$ for $i\ge 3$, so we can fix $a$, $r$ and $v_2$ to any value, while  all other higher ERE parameters are entirely determined by the full off-shell $T_\pi$ matrix and the latter three lowest terms of the ERE. Using the numerical values for $\delta_0$, $\delta_1$ and $\delta_2$  given in Eq.~\eqref{eq:delta13org}, in this case we obtain
\begin{eqnarray}
        a &=& -23.7588 \,{\rm fm}
        \nonumber \\
        r &=&  2.67286 \,{\rm fm}
        \nonumber \\
        v_2 &=& -0.57140 \,{\rm fm^3}
        \nonumber \\
        v_3 &=& 4.89257 \,{\rm fm^5}
        \nonumber \\
        v_4 &=& -29.4580 \,{\rm fm^7}
        \nonumber \\
        v_5 &=& 186.251 \,{\rm fm^9}
        \nonumber \\
        v_6 &=& -1228.10 \,{\rm fm^{11}} \label{eq:orden2infty}
\end{eqnarray}
This scheme cannot be reproduced by the $N/D_{22}$ solution, fixing $a$, $r$ and $v_2$, whose phase-shifts were shown in Fig.~\ref{fig:forder1} and ERE parameters given in Eq.~\eqref{eq:lowJuan}. The $N/D_{22}$ method provides different values of $v_{i\ge 3}$ than Eq.~\eqref{eq:orden2infty} since in the latter case $\delta_{i\ge 3}=0$, while in the former one $\delta_{i\ge 3}\neq 0$ (see Eq.~\eqref{eq:serie1}). 

\begin{figure}
\begin{center}
\includegraphics[width=7.5cm]{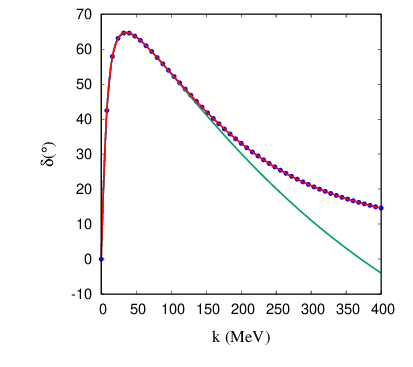}
\caption{\label{fig:deltao2inf} Phase-shifts from the ${\cal O} (Q^4)$ DR$_{\infty}$ method fixing
$a$, $r$ and $v_2$ (blue line with dots) compared with
the result of $N/D_{23}$ fixing also $v_3$ (red line). 
The green curve, corresponding to the  DR calculation at order 2 ($n=1$) discussed in Sec.~\ref{sec:DRorden2}, is also depicted for comparison.}
\end{center}
\end{figure}

The  $N/D_{23}$ calculation fixing also $v_3$ to 
the value in Eq.~\eqref{eq:orden2infty} reproduces the results of the ${\cal O} (Q^4)$ DR$_{\infty}$ scheme. The low energy parameters nicely agree and the phase-shifts are shown in Fig.~\ref{fig:deltao2inf}.

As a summary at order 4 ($n=2$) we get that there is a one parametric family of theories that fixes $a$, $r$ and $v_2$. If
one wants to use $v_3$ (or equivalently $\delta_3$) to fix $J_0^R$ there are two solutions so the theory is only
fixed when $v_4$ is also specified. This is to say five conditions in total, and we have found  that the $N/D_{33}$ solution agrees with the ${\cal O} (Q^4)$  DR calculation. On the other hand, it might look like  that one needs only three parameters to define the DR$_\infty$ calculation at this order. However, we obtained agreement considering $N/D_{23}$ fixing the values up to $v_3$ to guarantee that $\delta_3=0$, which means that the $N/D$ method needs to fix the four lowest energy constants.

\subsection{Order 6, $n=3$}
\label{order3}
At this order we have 4 renormalization conditions and 6 couplings so we can fix $\delta_0,\delta_1, \delta_2$ and $\delta_3$.

\subsubsection{${\cal O} (Q^6)$  DR scheme}
The couplings in terms of the renormalization constants and $J_0^R$ are given by
\begin{eqnarray}
	g_{00}(J_0^R) &=& -2 m_\pi^2 k_\pi^2 \left(g_{02}(J_0^R)-m_\pi^2g_{03}(J_0^R)\right) + \frac{1}{\delta_0+J_0^R}
        - \frac{(\delta_1+\delta_2k_\pi^2+\delta_3k_\pi^4)k_\pi^2}{(\delta_0+J_0^R)^2}
         \nonumber \\
       &&+ \frac{\delta_1(\delta_1+2\delta_2k_\pi^2)k_\pi^4}{(\delta_0+J_0^R)^3} - \frac{\delta_1^3k_\pi^6}{(\delta_0+J_0^R)^4}
        \nonumber \\
	g_{01}(J_0^R) &=& -\frac{10}{3} m_\pi^2 k_\pi^2 g_{03}(J_0^R)
        - \frac{\delta_1+2\delta_2k_\pi^2+\delta_3(3k_\pi^2-m_\pi^2)k_\pi^2}{2(J_0^R + \delta_0)^2}
        +\frac{\delta_1[\delta_1+ \delta_2(3k_\pi^2-m_\pi^2)] k_\pi^2}{(J_0^R + \delta_0)^3}\nonumber \\
       && - \frac{\delta_1^3k_\pi^2(3k_\pi^2-m_\pi^2)}{2(\delta_0+J_0^R)^4}
        \nonumber \\
	g_{11}(J_0^R) &=& -2 g_{02}(J_0^R)
        - \frac{\delta_2+3\delta_3k_\pi^2}{(J_0^R + \delta_0)^2}
        + \frac{\delta_1^2+6\delta_1\delta_2k_\pi^2}{(J_0^R + \delta_0)^3}
        - \frac{3\delta_1^3k_\pi^2}{(\delta_0+J_0^R)^4}
        \nonumber \\
	g_{12}(J_0^R) &=& -g_{03}(J_0^R)
        - \frac{\delta_3}{2(J_0^R + \delta_0)^2}
        + \frac{\delta_1\delta_2}{(J_0^R + \delta_0)^3}
        - \frac{\delta_1^3}{2(\delta_0+J_0^R)^4}
\end{eqnarray}
Here, we have a two-parametric ($g_{02}$ and $g_{03}$) family of theories that are equal on-shell and leads to 
the same LECs $\delta_{0,1,2,3}$ for any value of $J_0^R$. Now the next term is
\begin{eqnarray}
	\delta_4 &=& 
	\frac{\delta_2^2+ 2 \delta_1 \delta_3}{\delta_0+J_0^R}
	-\frac{3\delta_1^2 \delta_2}{(\delta_0+J_0^R)^2}
	+\frac{\delta_1^4}{(\delta_0+J_0^R)^3} \label{eq:delta4Orden3}
\end{eqnarray}
and in this way $(\delta_0+J_0^R)$ can be obtained from a cubic equation with coefficients determined by $\delta_1$, $\delta_2$, $\delta_3$ and $\delta_4$. Hence, we have 3 different solutions, not always real, for $J_0^R$.  The $\delta_5$ term will be different for each one, hence we expect that $\delta_5$ would need also to be fixed and maybe also $\delta_6$. So we proceed in the following way, we build a theory and check the $N/D_{44}$ and $N/D_{34}$ solutions. 

\begin{figure}
\begin{center}
\includegraphics[width=7.5cm]{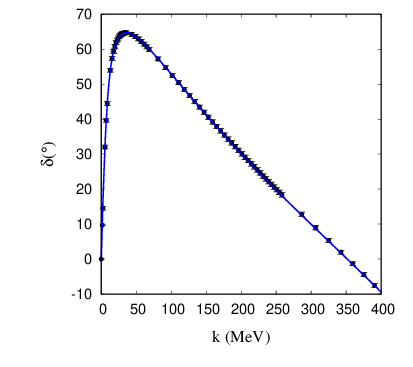}
\includegraphics[width=7.5cm]{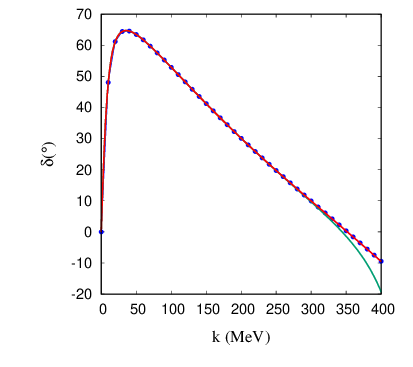}
\caption{\label{fig:deltao3}  Left: Comparison of the  $J_0^R$-finite ${\cal O} (Q^6)$   DR theory ($n=3$)  (blue line) with the Nijmegen phase-shifts (black dots) from Ref.~\cite{Stoks:1993tb}. Right: Comparison between the same  DR calculation (blue line with dots) with
the $N/D_{34}$ (green line) and $N/D_{44}$ (red line) phase-shifts. In the  DR scheme, we take  $\delta_0$, $\delta_1$ and $\delta_2$ from Eq. (\ref{eq:delta13org}), $\delta_3= -39.0717\,m^{-5}$ and $J_0^R/m=-2.34111$, which fixes $\delta_4=-93.2326\, m^{-7}$. In the $N/D_{44}$ ($N/D_{34}$) solution, the ERE parameters $a,\dots, v_6$ ($a,\dots, v_5$) observables are set to the values obtained in the  DR scheme and given in Eq.~\eqref{eq:orden3}.}
\end{center}
\end{figure}

We fix $\delta_0$, $\delta_1$ and $\delta_2$ to the values of Eq. (\ref{eq:delta13org}) and take $\delta_3=-39.0717\,m^{-5}$. Then we vary the value of $J_0^R$ to improve the description of the Nijmegen phase-shifts at high energies. In this way, we fix $J_0^R = -2.34111\, m$ and the couplings are now given by
\begin{eqnarray}
	\left[g_{00}+ 2  m_\pi^2 k_\pi^2 (g_{02}-g_{03} m_\pi^2)\right] m &=& -0.36302
        \nonumber \\
	\left(g_{01}+ 10  m_\pi^2 k_\pi^2 g_{03}/3 \right) m^3 &=&  0.59804
        \nonumber \\
	\left(g_{11}+2  g_{02}\right) m^5 &=&  -0.94050
        \nonumber \\
	(g_{12} +g_{03}) m^7 &=&  0.42551
\end{eqnarray}
These values lead to $\delta_4=-93.2326\, m^{-7}$.
The ERE parameters in the  DR turn out to be   
\begin{eqnarray}
        a &=& -23.7588 \,{\rm fm}
        \nonumber \\
        r &=&  2.67286 \,{\rm fm}
        \nonumber \\
        v_2 &=& -0.571396 \,{\rm fm^3}
        \nonumber \\
        v_3 &=& 5.00024 \,{\rm fm^5}
        \nonumber \\
        v_4 &=& -29.2850 \,{\rm fm^7}
        \nonumber \\
        v_5 &=&  185.605\,{\rm fm^9}
        \nonumber \\
        v_6 &=& -1224.71 \,{\rm fm^{11}}
        \nonumber \\
        v_7 &=& 8331.90 \,{\rm fm^{13}}
        \nonumber \\
        v_8 &=& -58015.9 \,{\rm fm^{15}}
        \nonumber \\
        v_9 &=& 411403 \,{\rm fm^{17}} \label{eq:orden3}
\end{eqnarray}
and find the same values from the $N/D_{44}$ solution. The phase-shit is displayed in Fig.~\ref{fig:deltao3}. 
In the left panel, we compare the  DR calculation (blue line) with the  data from~\cite{Stoks:1993tb}. On the right, the blue line with dots shows again the  DR calculation, which perfectly agrees with the $N/D_{44}$ solution (red curve), but not with the $N/D_{34}$ solution (green curve).

\subsubsection{${\cal O} (Q^6)$ DR$_\infty$ scheme}

\begin{figure}
\begin{center}
\includegraphics[width=7.5cm]{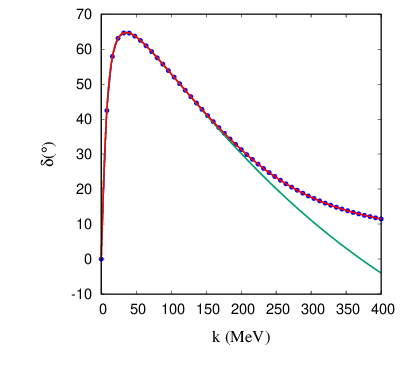}
\caption{\label{fig:deltao3inf}  Comparison of the ${\cal O} (Q^6)$ ($n=3$)   DR$_\infty$ theory (blue line with dots) with the $N/D_{34}$ (red line) phase-shifts. In the DR$_\infty$ scheme, we take  $\delta_0$, $\delta_1$ and $\delta_2$ from Eq. (\ref{eq:delta13org}) and $\delta_3= -39.0717\,m^{-5}$. In  the $N/D_{34}$ calculation, the constraints from setting $\delta_4$ and $\delta_5$ to zero need also to be taken into account. Note that the $N/D_{34}$ phase-shifts shown here differ from that  displayed in Fig.~\ref{fig:deltao3}, where $\delta_4$ and $\delta_5$ are fixed to reproduce $v_4$ and $v_5$ in Eq. (\ref{eq:delta13org}). For comparison, the green curve shows the original finite-$J_0^R$  DR calculation of Ref.~\cite{Nieves:2003uu} at order ${\cal O} (Q^2)$  ($n=1$) which produces the ERE parameters collected in Eq.~\eqref{eq:lowJuan}.}
\end{center}
\end{figure}

In the case $J_0^R\to\infty$, following a procedure similar to those outlined in Subsecs.~\ref{sec:drinftyorden1} and \ref{sec:drinftyorden2}, we find that the $N/D_{34}$ solution agrees with the ${\cal O} (Q^6)$ DR$_{\infty}$ result. The phase-shifts are compared in Fig.~\ref{fig:deltao3inf}. In the ${\cal O} (Q^6)$ DR$_{\infty}$ scheme, all $\delta_{i\ge 4}$ are set to zero, while as discussed in previous sections, when $J_0^R$ is kept finite $\delta_4$ could be adjusted to any value by tuning $J_0^R$. The higher $\delta_{i\ge 5}$ observables are not zero, and all of them are correlated since they become functions of the LECs $\delta_0,\delta_1,\delta_2,\delta_3$ and $\delta_4$.

\begin{table}
\begin{center}
\begin{tabular}{lcccc}
\hline
\hline
\multicolumn{5}{c}{finite $J_0^R$} \\
\hline
                      &${\cal O}(Q^0)$ & ${\cal O}(Q^2)$ & ${\cal O}(Q^4)$ &${\cal O}(Q^6)$ \\
$n$ in  DR           & 0          & 1          & 2          & 3 \\
parameters  in  DR      & 2          & 3          & 5          & 7 \\
independent parameters in  DR  [$\delta_i$] & 1          & 3          & 4          & 5 \\
$N/D$ calculation       & $N/D_{11}$ & $N/D_{22}$ & $N/D_{33}$ & $N/D_{44}$  \\
parameters in $N/D$     & 1          & 3          & 5          & 7 \\
\hline
\hline
\multicolumn{5}{c}{$J_0^R\to\infty$} \\
\hline
$n$ in DR$_{\infty}$           & 0          & 1          & 2          & 3 \\
independent parameters  in DR$_{\infty}$  [$\delta_i$] & 1          & 2          & 3          & 4 \\
$N/D$ calculation       & $N/D_{11}$ & $N/D_{12}$ & $N/D_{23}$ & $N/D_{34}$ \\
parameters in $N/D$     & 1          & 2          & 4          & 6 \\
\hline
\hline
\end{tabular}
\caption{\label{sum} Summary of the equivalence of the  DR calculation up to ${\cal O} (Q^6)$ ($n=3$) with the $N/D_{nd}$ solutions. The number of LECs fixed in the $N/D_{nd}$ solution is $n+d-1$, while the parameters in the  DR scheme are those appearing in the bare $V_s$ potential plus $J_0^R$.}
\end{center}
\end{table}

We provide in Table~\ref{sum} a summary of the equivalence between the  DR and DR$_\infty$  calculations up to ${\cal O} (Q^6)$  ($n=3$) with the exact $N/D$ method.

\subsection{${\cal O} (Q^8)$ DR$_\infty$ scheme (order 8, $n=4$)}
\begin{figure}
\begin{center}
\includegraphics[width=9cm]{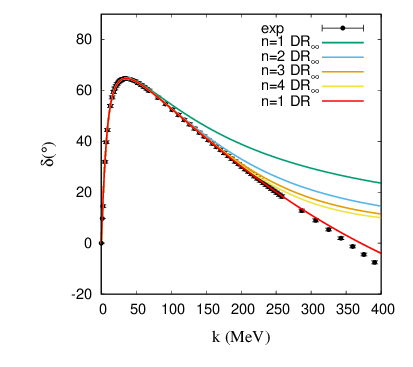}
\caption{\label{fig:deltaO4inf} Comparison of the ${\cal O} (Q^2)$, ${\cal O} (Q^4)$, ${\cal O} (Q^6)$ and ${\cal O} (Q^8)$  DR$_\infty$ phase-shifts, obtained using the numerical values for $\delta_{0,1,2,3,4}$ given in Eqs.~\eqref{eq:delta012} and \eqref{eq:delta34}. The corresponding ERE parameters from each calculation are compiled in  Table \ref{tab:deltaO4inf}. The red curve stands for the original finite-$J_0^R$  DR calculation of Ref.~\cite{Nieves:2003uu} at order ${\cal O} (Q^2)$, which produces the ERE parameters collected in the last column of Table \ref{tab:deltaO4inf}. The black dots with error bars are the experimental phase-shifts from Ref.~\cite{Stoks:1993tb}.  }
\end{center}
\end{figure}
\begin{table}
\begin{tabular}{lrrrrr}
\hline
\hline
 DR$_\infty$& ${\cal O} (Q^2)$ & ${\cal O} (Q^4)$ & ${\cal O} (Q^6)$ & ${\cal O} (Q^8)$ & Ref.~\cite{Nieves:2003uu} \\
\hline
\hline
$a$ [fm]          & $-$23.7588  & $-$23.7588  & $-$23.7588  & $-$23.7588  & $-$23.7588 \\
$r$ [fm]          &  $\underline{2.67286}$ &  2.67286  &  2.67286  &  2.67286  &  2.67286  \\
$v_2$ [fm$^3$]    & $-$0.838491&  $-\underline{0.571396}$ &  $-$0.571396 & $-$0.571396 & $-$0.571396 \\
$v_3$ [fm$^5$]    &  4.57626  &  4.89256  &     $\underline{5.00024}$  & 5.00024  & 5.00024  \\
$v_4$ [fm$^7$]    & $-$27.7418  & $-$29.4580  & $-$29.3304  & $-\underline{29.2870}$  & $-$29.2870  \\
$v_5$ [fm$^9$]    &  177.297  &  186.250  &  185.532  &  185.584  &  185.601  \\
$v_6$ [fm$^{11}$] & $-$1176.81  & $-$1228.10  & $-$1224.44  & $-$1224.73  & $-$1224.71  \\
$v_7$ [fm$^{13}$] &  8037.07  &  8351.25  &  8330.48  &  8331.95  &  8331.84  \\
$v_8$ [fm$^{15}$] & $-$56116.2  & $-$58134.6  & $-$58007.7  & $-$58016.1  & $-$58015.5  \\
$v_9$ [fm$^{17}$] &  398738   &  412166   &  411353   &  411404   &  411401   \\
\hline
\hline
\end{tabular}
\caption{\label{tab:deltaO4inf} First ten parameters of the ERE from the ${\cal O} (Q^2)$, ${\cal O} (Q^4)$, ${\cal O} (Q^6)$ and ${\cal O} (Q^8)$  DR$_\infty$ calculations displayed in Fig.~\ref{fig:deltaO4inf} and of  the ${\cal O} (Q^2)$ finite-$J_0^R$   DR calculation of Ref.~\cite{Nieves:2003uu}.  At each order, we underline  the last ERE parameter that is used as input.}
\end{table}
Finally, we discuss some results obtained for $n=4$ [${\cal O} (Q^8)$] within the  DR$_\infty$ scheme to illustrate that it provides a systematic and improvable description of any ERE. We consider to reproduce the phase-shifts obtained in the ${\cal O}(Q^2)-$DR calculation, derived in Ref.~\cite{Nieves:2003uu} and discussed here in Subsec.~\ref{sec:DRorden2}, and shown\footnote{As seen in the plots of  Fig.~\ref{fig:forder1}, the $N/D_{22}$ theory with $a,r$ and $v_2$ as input reproduces the ${\cal O}(Q^2)-$DR calculation.} in Fig.~\ref{fig:forder1}. The ${\cal O} (Q^8)$ DR$_\infty$ scheme fixes $\delta_0,\delta_1,\delta_2,\delta_3$ and $\delta_4$, which for the finite-$J_0^R$ ${\cal O}(Q^2)-$DR calculation  of Ref.~\cite{Nieves:2003uu} are
\begin{equation}
	\delta_0 = -0.08111\, m\,, \quad
	\delta_1  = -7.50329\, m^{-1}\,, \quad
	\delta_2  = -17.1221\, m^{-3} \label{eq:delta012}
\end{equation}
given in Eq.~\eqref{eq:delta13org}, and $\delta_3$ and $\delta_4$ determined from the correlation of Eq.~\eqref{eq:serie1}
\begin{equation}
   \delta_3  = \frac{\delta_2^2}{\delta_1} = -39.0717\, m^{-5}\,, \quad \delta_4  = \frac{\delta_2^3}{\delta_1^2}  =-89.1596\, m^{-7}\,.\label{eq:delta34}
\end{equation}
The higher $\delta_{i\ge 5}$ observables are set to zero at this order. The phase-shifts are shown\footnote{As expected, the  $N/D_{45}$ theory exactly reproduces these phase-shifts, using $a,r,v_2,v_3,v_4,v_5,v_6$ and $v_7$ from the ERE of the ${\cal O} (Q^8)$  DR$_\infty$ results to fix the eight subtractions of this $N/D$ solution.} in Fig.~\ref{fig:deltaO4inf} and the first ten parameters of the ERE for this calculation are collected in  Table \ref{tab:deltaO4inf}. To quantify the convergence of the DR$_\infty$ approach, both in the figure and the table, we also show results for  the lower-order calculations and for the original finite-$J_0^R$  DR calculation of Ref.~\cite{Nieves:2003uu} at order ${\cal O} (Q^2)$. We observe that the DR$_\infty$ renormalization progressively improves the description order by order and slowly tends to reproduce the given theory. Furthermore, we have verified that the effective range approximation, taking as input the same ERE parameters used in the corresponding ${\cal O} (Q^{2n})$ DR$_\infty$ solution, performs noticeably worse.

We also note that the higher-order-term correlation $\delta_{m\ge 3}=\delta_2^{m-1}/\delta_1^{m-2}$ [Eq.~\eqref{eq:serie1}] induced in the finite-$J_0^R$ ${\cal O}(Q^2)-$DR calculation turns out to be very  successful phenomenologically in this case.

\section{Proof of equivalence between DR-LSE and $N/D$ methods}
\label{sec:proof}
\subsection{The LHC in the DR-LSE amplitude}
Along this work we have studied in detail the application of the exact $N/D$ method and the LSE, solved in DR, for the case of  a regular finite-range potential $V_\pi$, and compare the predictions of the two approaches. In this section we will demonstrate that the exact $N/D$ method, mathematically expressed as an $N/D_{nd}$ integral equation with an {\it adequate} choice for the number of subtractions $n$ and $d$, is equivalent to the  DR calculation. For that we start from Eq.~\eqref{eq:tsol}. If we make an analytical continuation to the complex $k$ plane, we see that the LHC of $T(k)$ is given by the LHC of $T_\pi(k)$ plus the LHC of the second term in the right hand side of Eq.~\eqref{eq:tsol}. If this second term did not have  LHC then $T(k)$ and $T_\pi(k)$ would have the same LHC. This is the key point, since it means that to obtain the $T(k)$ with the $N/D$ method we would only need the discontinuity $\Delta(A)$ of the  OPE $T_\pi(k)$ matrix on the LHC. 

The limit $\sqrt{-A}\to \infty$ of $\Delta(A)$ for the OPE was given in Ref.~\cite{Entem:2016ipb}\footnote{This is the asymptotic behavior of the discontinuity of $-2\pi^2 T_\pi/M_N$ on the LHC, with the $-2\pi^2/M_N$ factor accounting for the different convention used here and in \cite{Entem:2016ipb} to relate the $T$ and $S$ matrices.}
\begin{eqnarray}
\lim_{\sqrt{-A}\to \infty}	\Delta(A) &=& \frac{\lambda \pi^2}{M_N A} e^{\frac{2\lambda}{\sqrt{-A}} {\rm arctanh}\big(1-\frac{m_\pi}{\sqrt{-A}}\big)} 
\end{eqnarray}
with $\lambda=\alpha_\pi M_N$. The function $\Delta(A)$ decreases faster than $(-A)^{-1/2}$ and it was shown 
in~\cite{Guo:2013rpa} that then the $N/D$ method with multiple subtractions has always a solution and it is unique, using
the  Fredholm theorem \cite{TricomiIE}. This means that if we
build a theory for $T(k)$ that fixes a certain number of LECs, it will be the same as that derived from 
the $N/D$ method fixing these LECs, although the specific number of subtractions will be discussed in detail below in the next subsection. This is because, it will be the unique solution that has the same left- and right-hand cuts and the same fixed LECs. If $T(k)$ were different the $N/D$ method would have two solutions.

Now we focus on the second term in the right-hand side of Eq.~\eqref{eq:tsol}. It includes the pion-distorted two-nucleon loop function $J_0(k)$ given in Eq.~\eqref{eq:j0}. Clearly $J_0(k)$ does not have LHC, since the fully off-shell  $T_\pi(q',q; k^2)$ matrix does not have it for complex $k$ and real integration variables $q$ and $q'$, and 
$J_0(k)$ only has the RHC.

The other integral that appears is $L_0(k)$ defined in Eq.~\eqref{eq:l0} involves the half off-shell OPE matrix.  We have
\begin{eqnarray}
	L_0(i\bar k+\epsilon) &=&
	\lim_{\epsilon\to 0} 
	\int_0^\infty dq \frac{q^2 T_\pi\left(i\bar k+\epsilon,q; (i\bar k+\epsilon)^2\right)}{-\bar k^2-q^2+i\epsilon}
\end{eqnarray}
with $k=i\bar k +\epsilon$ and $\bar k \in  \mathbb{R}$. We are interested in
\begin{eqnarray}
	\Delta L_0(\bar k) &=& \lim_{\epsilon\to 0} [ L_0(i\bar k+\epsilon) - L_0(i\bar k-\epsilon) ]
\end{eqnarray}
Next, we will formally show that $\Delta L_0(\bar k)=0$, which essentially follows from the fact that the $V_\pi(i\bar k,q)$ is real for $q\ge0$ and that the dynamical cuts of $T_\pi(k,q;k^2)$ stem from those of the potential~\cite{Oller:2018zts}. The latter statement implies that the cuts of $L_0(i\bar k+\epsilon)$ are given by
\begin{eqnarray}
	q(x) &=& i\bar k + \epsilon \pm i \sqrt{m_\pi^2+x^2}
	\nonumber \\
	q(x) &=& -i\bar k - \epsilon \pm i \sqrt{m_\pi^2+x^2}
\end{eqnarray}
with $x\in {\mathbb R}$. Thus,  we see that for $L_0(i\bar k+\epsilon)$, the $q-$integration path does not cross the second set of cuts, but it crosses one of the cuts of the first set.
It happens when
\begin{eqnarray}
	q &=& i\bar k + \epsilon - i (m_\pi+\nu) = \epsilon
\end{eqnarray}
So $\bar k - m_\pi = \nu >0$ and the cut starts at $\nu =0$ ($x=0$). The deformed contour that avoid the cut is
\begin{eqnarray}
	q &=& \epsilon - \delta + i \nu 
\end{eqnarray}
with $\nu$ from 0 to $\bar k-m_\pi$ going up and
\begin{eqnarray}
	q &=& \epsilon + \delta + i \nu 
\end{eqnarray}
with $\nu$ from $\bar k-m_\pi$ to 0 going down. 

For the case $L_0(i\bar k-\epsilon)$ the cuts are
\begin{eqnarray}
	q(x) &=& i\bar k - \epsilon \pm i \sqrt{m_\pi^2+x^2}
	\nonumber \\
	q(x) &=& -i\bar k + \epsilon \pm i \sqrt{m_\pi^2+x^2}
\end{eqnarray}
and now one of the  cuts of the second set is crossed. Again we deform the contour of integration for
\begin{eqnarray}
	q &=& \epsilon - \delta - i \nu 
\end{eqnarray}
with $\nu$ from 0 to $\bar k-m_\pi$ going down and
\begin{eqnarray}
	q &=& \epsilon + \delta - i \nu 
\end{eqnarray}
with $\nu$ from $\bar k-m_\pi$ to 0 going up. 

The rest of the integration path is the same in both cases, and hence
\begin{eqnarray}
	-i\Delta L_0(\bar k )&=& -i\lim_{\epsilon\to 0} ( L_0(i\bar k+\epsilon) - L_0(i\bar k-\epsilon) )
	\nonumber \\ &=&
	\lim_{\epsilon\to 0}\lim_{\delta\to 0} \Bigg\{
	\int_0^{\bar k-m_\pi} d\nu \frac{\nu^2 T_\pi\left[i\bar k+\epsilon,\epsilon - \delta + i \nu;(i\bar k+\epsilon)^2\right]}{\bar k^2-\nu^2}
	+\int^0_{\bar k-m_\pi} d\nu \frac{\nu^2 T_\pi\left[i\bar k+\epsilon,\epsilon + \delta + i \nu;(i\bar k+\epsilon)^2\right]}{\bar k^2-\nu^2}
	\nonumber \\ &+&
	\int_0^{\bar k-m_\pi} d\nu \frac{\nu^2 T_\pi\left[i\bar k-\epsilon,\epsilon-\delta-i\nu;(i\bar k-\epsilon)^2\right]}{\bar k^2-\nu^2}
	+\int^0_{\bar k-m_\pi} d\nu \frac{\nu^2 T_\pi\left[i\bar k-\epsilon,\epsilon+\delta-i\nu;(i\bar k-\epsilon)^2\right]}{\bar k^2-\nu^2}
	\Bigg\}
	\nonumber \\ &=&
	\int_0^{\bar k-m_\pi} d\nu \frac{\nu^2}{\bar k^2-\nu^2} 
	\lim_{\epsilon\to 0}\lim_{\delta\to 0} \bigg\{
	 T_\pi\left[i\bar k+\epsilon,\epsilon - \delta + i \nu;(i\bar k +\epsilon)^2\right]
 	-T_\pi\left[i\bar k+\epsilon,\epsilon + \delta + i \nu;(i\bar k +\epsilon)^2\right] \bigg\}
	\nonumber \\ &+&
	\int_0^{\bar k-m_\pi} d\nu \frac{\nu^2}{\bar k^2-\nu^2} 
	\lim_{\epsilon\to 0}\lim_{\delta\to 0} \bigg\{
	T_\pi\left[i\bar k-\epsilon,\epsilon-\delta-i\nu;(i\bar k -\epsilon)^2\right]
	-T_\pi\left[i\bar k-\epsilon,\epsilon+\delta-i\nu;(i\bar k -\epsilon)^2\right] \bigg\}
\end{eqnarray}
Now we define
\begin{eqnarray}
	2i \Delta T_\pi(\nu,\bar k) &=& 
	\lim_{\epsilon\to 0}\lim_{\delta\to 0} \bigg\{
	T_\pi\left[i\bar k+\epsilon,\epsilon - \delta + i \nu;(i\bar k +\epsilon)^2\right]
-T_\pi\left[i\bar k+\epsilon,\epsilon + \delta + i \nu;(i\bar k +\epsilon)^2\right] \bigg\}
\end{eqnarray}
with the discontinuity being purely imaginary as demonstrated in Ref.~\cite{Oller:2018zts}.
Compared with Eq.~\eqref{250522.1}, the order of the arguments is reversed but this does not matter since~\cite{Oller:2018zts}
\begin{eqnarray}
	T_\pi(p',p;k^2) &=& T_\pi(p,p';k^2)\,.
\end{eqnarray}
 On the other hand, we make use of the relations (Eqs.~(5.10) y (5.12) of Ref.~\cite{Oller:2018zts} which are demonstrated in Appendix C of that reference)
\begin{eqnarray}
	T_\pi(p',k;k^2) &=& T_\pi^*(p'^*,k^*;k^{*2})
	\\
	T_\pi(p',k;k^2) &=& T_\pi(-p',k;k^2)
\end{eqnarray}
and it follows
\begin{eqnarray}
	T_\pi\left[i\bar k-\epsilon,\epsilon-\delta-i\nu;(i\bar k -\epsilon)^2\right] 
	-T_\pi\left[i\bar k-\epsilon,\epsilon+\delta-i\nu;(i\bar k -\epsilon)^2\right] &=&
	\nonumber \\
	T_\pi^*\left[-i\bar k-\epsilon,\epsilon-\delta+i\nu;(-i\bar k -\epsilon)^2\right] 
	-T_\pi^*\left[-i\bar k-\epsilon,\epsilon+\delta+i\nu;(-i\bar k -\epsilon)^2\right] &=&
	\nonumber \\
	T^*_\pi\left[i\bar k+\epsilon,\epsilon-\delta+i\nu;(i\bar k +\epsilon)^2\right] -
	 T^*_\pi\left[i\bar k+\epsilon,\epsilon+\delta+i\nu;(i\bar k +\epsilon)^2\right] 
	&=& -2i \Delta T_\pi(\nu,\bar k)
\end{eqnarray}
Thus, we finally get
\begin{eqnarray}
	\Delta L_0(\bar k) &=& 
	-i\int_0^{\bar k-m_\pi} d\nu \frac{\nu^2}{\nu^2-\bar k^2} 
	2i \Delta T_\pi(\nu,\bar k)
	-i\int_0^{\bar k-m_\pi} d\nu \frac{\nu^2}{\nu^2-\bar k^2} 
	2(-i) \Delta T_\pi(\nu,\bar k)
\end{eqnarray}
and therefore
\begin{eqnarray}
	\Delta L_0(\bar k) &=& 
	\int_0^{\bar k-m_\pi} d\nu \frac{\nu^2}{\nu^2-\bar k^2} 
	2 \Delta T_\pi(\nu,\bar k)
	-\int_0^{\bar k-m_\pi} d\nu \frac{\nu^2}{\nu^2-\bar k^2} 
	2 \Delta T_\pi(\nu,\bar k) = 0
\end{eqnarray}
This proofs that $L_0(k)$ has no LHC and the LHC discontinuity of $T$ is equal to the LHC discontinuity of the OPE $T_\pi$ amplitude.

\subsection{Independent LECs in the DR-LSE amplitude and subtractions in the $N/D_{nd}$ method}
\label{sec:numerosparam}
Now we discuss  how many LECs $\delta_i$ are needed to fix the DR-LSE theory. Let us suppose that we are working at order ${\cal O}(Q^{2n})$, then the renormalized contact potential in Eq.~\eqref{eq:tsol} is  $\widetilde V_s(k)=\sum_{i=0}^n h_i k^{2i}$ and the LECs $\delta_i$ are introduced by
\begin{eqnarray}
	\frac{1}{\sum_{i=0}^n h_i k^{2i}} - J_0^R &=& \sum_{i=0}^\infty \delta_i k^{2i}\nonumber\\
    d_0&=&\frac{1}{h_0}= \delta_0 + J_0^R \nonumber\\
	d_i &=& \delta_i=-\frac{1}{h_0}  \sum_{m=1}^{{\rm min}(i,n)} h_m d_{i-m},\quad i>0
\end{eqnarray}
The LEC $\delta_0$ and $1/h_0$ determines $J_0^R$ and looking at $i=1,\ldots,n+1$,  we find  a set of non-linear  equations with unknowns $h_{i=1\cdots n}/h_0$ and $1/h_0$
\begin{eqnarray}
	\delta_1 &=& -\frac{h_1}{h_0} (\delta_0+J_0^R)
	\nonumber \\
	\delta_2 &=& -\frac{h_1}{h_0} \delta_1 -\frac{h_2}{h_0} (\delta_0+J_0^R)
	\nonumber \\ &\vdots&
	\nonumber \\ 
	\delta_n &=& \left(-\sum_{m=1}^{n-1} \frac{h_m}{h_0} \delta_{n-m}\right) -\frac{h_n}{h_0} (\delta_{0} +J_0^R)
	=-\left(\frac{h_1}{h_0}\delta_{n-1}+\ldots+\frac{h_n}{h_0}(\delta_{0}+J_0^R)\right) \nonumber \\ 
   \delta_{n+1} &=& - \sum_{m=1}^{n}\frac{h_m}{h_0}  \delta_{n+1-m} =
	-\left(\frac{h_1}{h_0} \delta_{n} + \ldots + \frac{h_n}{h_0} \delta_{1}\right)  
\end{eqnarray}
which might have multiple solutions. However, considering $i=n+1,\ldots,2n$
\begin{eqnarray}
	\delta_{n+1} &=& -\frac{1}{h_0} \sum_{m=1}^{n} h_m \delta_{n+1-m} =
	-\left(\frac{h_1}{h_0} \delta_{n} + \ldots + \frac{h_n}{h_0} \delta_{1}\right) 
	\nonumber \\ &\vdots&
	\nonumber \\ 
	\delta_{2n} &=& -\frac{1}{h_0} \sum_{m=1}^{n} h_m \delta_{2n-m} =
	-\left(\frac{h_1}{h_0} \delta_{2n-1} + \ldots + \frac{h_n}{h_0} \delta_{n}\right) 
	\label{eq:hie}
\end{eqnarray}
one finds a system of $n$ equations with $n$ unknowns, namely $h_{i=1\cdots n}/h_0$, that involves the LECs
$\delta_1, \delta_2,\ldots, \delta_{2n}$ and has a well defined solution. In addition,  $h_0$ and $J_0^R$ can be determined from 
\begin{eqnarray}
	h_0 &=& -\frac{h_1}{h_0} \frac{1}{\delta_1}
	\nonumber \\
	J_0^R &=& \frac{1}{h_0} - \delta_0 \label{eq:hie-0}
\end{eqnarray}
This proofs that the $(2n+1)$ $\delta_i$ values define always unambiguously the DR-LSE series. As shown before this is equivalent to fix the $2n+1$ lowest ERE parameters which corresponds to $N/D_{n+1n+1}$. However, this does not proof that in some cases a lower number of $\delta_i$ could  not be sufficient to fix the theory.

However, it is clear that $\delta_0,\cdots \delta_{n+1}$ should be insufficient to fix the DR-LSE series due to the correlation between the higher $\delta_{i\ge n+2}$ LECs,  induced by the renormalization procedure when $J_0^R$ is kept finite. The question is how many of these ERE-like coefficients are additionally needed to unequivocally fix the theory. The answer should be related to the rank of the non-linear (polynomial) equation employed to obtain $J_0^R$.  At order ${\cal O}(Q^{4})$, it was a second order equation [Eq.~\eqref{eq:delta3Orden2}], which might have two real solutions for  $J_0^R$  [Eq.~\eqref{eq:J0R2}], with each of them inducing a different correlation between the higher order $\delta_{i\ge 4}$ terms [Eq.~\eqref{eq:seriesorden2}]. At this order and to completely disentangle the DR-LSE series, the additional LEC $\delta_4$ was specified (see Fig.~\ref{fig:deltao2}). Thus, we ended up with the $N/D_{33}$ solution, which implements five subtractions fixing  $\delta_0$, $\delta_1$, $\delta_2$, $\delta_3$ and $\delta_4$.   At the next order ${\cal O}(Q^{6})$, the subtraction constant $J_0^R$ is determined from a cubic  equation [Eq.~\eqref{eq:delta4Orden3}] and thus three different real solutions might exist, which could lead to different correlations for the $\delta_{i\ge 5}$ terms. Hence, two additional LECs  $\delta_5$  and  $\delta_6$ are needed to fully disentangle the series, and the correspondence to  the $N/D_{44}$ solution, which fixes seven ERE parameters, is apparent. In general at  ${\cal O}(Q^{2n})$, $J_0^R$ will be obtained from a $n-$rank nonlinear equation, which might have $n$ real solutions for $J_0^R$, and therefore lead to $n$ different DR-LSE series. To fully specify the DR-LSE theory, in addition  to $\delta_{0,\cdots,n+1}$ that might be used to fix $J_0^R$ and the renormalized contact potential at this order, the next $(n-1)$ $\delta_{n+2,\cdots 2n}$ LECs are required to select a particular correlation between the higher order terms. This is a total of $(n+2)+(n-1)=2n+1$ LECs in accordance to the discussion of upper panel of Table~\ref{sum} and Eqs.~\eqref{eq:hie}-\eqref{eq:hie-0}. Besides, the correspondence with  the $N/D_{n+1n+1}$ theory is straightforward.

If the set of $\delta_{0,\cdots,n+1}$ LECs does not provide $n$ different real values for $J_0^R$, and some solutions are complex, there will be also a smaller number of different possible DR-series.  In that case, it might happen that specifying less than $2n+1$ LECs ($\delta_i$)  is enough to specify the theory.

 Consider there exists an $N/D_{n'd'}$ solution  requiring a lower number of subtractions (i.e., $n'+d' < 2n+2$) that fixes the $n'+d'-1$ first ERE parameters and it is equivalent to the  DR calculation. Then  the next $2n+1-(n'+d'-1)$ higher ERE parameters will be those of the  DR calculation, and therefore obtaining the $N/D_{n+1,n+1}$ with these ERE parameters will generate the same $N/D_{n'd'}$ amplitude. Hence, the proposed solution will be always equivalent to the  DR calculation.

In the case $J_0^R \to \infty$, we have that $h_0 \to 0$ and
\begin{eqnarray}
\label{250523.1}
	\delta_0 &=& \frac{1}{h_0} - J_0^R \,
	\nonumber \\
    \frac{h_i}{h_0} &\to& 0,\, \quad i=1\cdots n  \,\Rightarrow
    \delta_i \to 0 \quad \forall\, i \ge n+1\nonumber \\
    \frac{h_i}{h_0^2} &\to & -\delta_i,\,  \quad i=1\cdots n 
\end{eqnarray}
  We need $2n$ LECs ($\delta_{i=0,1,\cdots 2n-1}$) values to define always unambiguously the series in this case. The first $n$ LECs fix $h_{i=1\cdots n}/h_0^2 \to -\delta_i$, while the next  $(n-1)$ conditions, $\delta_{i=n+1\cdots 2n-1}\to 0$,  guarantee\footnote{ The equation used to determine $J_0^R$ from the $\delta_{n+1}$ LEC takes the form $\delta_{n+1}(J_0^R + \delta_0)^n + \ldots$, where the ellipsis denotes a polynomial of degree $(n-1)$ in $J_0^R$ (see for instance Eqs.~\eqref{eq:delta3Orden2} or \eqref{eq:delta4Orden3} for the $n = 2$ and $n=3$ cases, respectively).  In the limit $J_0^R \to \infty$, it follows that $\delta_{n+1} [J_0^R]^n$ scales as $[J_0^R]^{n-1}$, as consequence of Eq.~\eqref{250523.1} and that
$\displaystyle{\delta_{n+1} \to \frac{2h_n h_1}{h_0^3} + \ldots,}$ where the ellipsis indicates other possible quadratic terms in the $h_i$, also divided by $h_0^3$.
Therefore, in the asymptotic regime $J_0^R \to \infty$, the leading behavior of the equation becomes that of a polynomial of degree $(n-1)$, rather than degree $n$ as in the finite $J_0^R$ case. Consequently, by the same reasoning as applied for the finite case, one would expect that in the large $J_0^R$ limit additional $(n - 1)$ coefficients $\delta_i$ (with $i = n+1, \ldots, 2n-1$) must be specified. }  that $h_0\to 0$, and hence $h_{i=1\cdots n}/h_0 \to 0$. Additionally, one needs  $\delta_0$ to fix the departure between $1/h_0$ and the UV subtraction constant $J_0^R$, when the latter grows and  becomes sufficiently large,
\begin{equation}
\frac{1}{h_0} - J_0^R= \delta_0 
\end{equation}

\section{Conclusions.}
\label{conclusions}
We have solved the LSE with a two-particle irreducible amplitude that consists of a regular finite-range potential and additional contact terms with derivatives. We have followed the scheme derived in Ref.~\cite{Nieves:2003uu}, based on  distorted wave theory and dimensional regularization, and have analyzed the spin singlet nucleon-nucleon $S-$wave as case of study. In this partial wave, we have considered  the regular one-pion exchange (OPE) potential  and up to ${\cal O}(Q^6)$ (six derivatives) contact interactions. We have shown that the scattering amplitude solution of the LSE fulfills exact elastic unitarity, this is to say contains the appropriate RHC, and inherits the LHC of the long-distance OPE amplitude. 

We have distinguished two different renormalization schemes, by taking the loop function, in presence of the long-range interaction, at threshold ($J^R_0$) finite or infinite. Both of them lead to renormalized amplitudes that differ only in higher-order terms than those considered in the short-range contact potential. 

Furthermore, we have proven that the LSE amplitude coincides with that obtained from the exact $N/D_{nd}$ calculation, with the appropriate number and typology of subtractions to reproduce the effective range parameters taken as input to renormalize the LSE amplitude. 

 The generalization of these results to a  higher  number of derivatives is straightforward.

Concerning the DR-LSE approach, there exist two different strategies to systematically improve the reliability/accuracy of Eq.~\eqref{eq:tsol}. Given a long-distance part of the amplitude ($T_\pi$ in this work),  one can increase the order of the short-distance potential, as we have explored in this paper.  In parallel, one might also improve the long-distance interaction and add the two-pion exchange (TPE) potential.  This seems physically appropriate when trying to describe phase-shifts for momenta above $k\gtrsim m_\pi$.  In general, a smaller number of contact-terms should be needed to reproduce the physical data when the long-distance amplitude includes not only OPE, but also TPE. At low momenta, the explicit inclusion of TPE is usually not necessary and its effects can be efficiently taken into account by means of a reduced number of  counterterms. But as the momentum increases, describing the phase-shifts from the OPE alone should be less efficient and a higher order  in $V_s$ should be needed. This is illustrated here in Fig.~\ref{fig:deltaO4inf}, where it is shown that the OPE-DR$_\infty$ scheme, although progressively improving order by order the phase-shift above $k=m_\pi$ , converges very slowly\footnote{We have no explanation of why the original finite-$J_0^R$  DR calculation of Ref.~\cite{Nieves:2003uu} at order ${\cal O} (Q^2)$ (red curve in the figure), with only three undetermined LECs and whose long-distance part includes only OPE, works so well and gives a reasonable reproduction of the Nijmegen phase-shifts up to almost $k\sim$ 300 MeV. For these momenta, we think that explicit inclusion of the TPE potential would be necessary and we believe that the good performance of the finite DR calculation ${\cal O} (Q^2)$ $J_0^R$ could be a mere coincidence.}. In the $N/D$ method, with multiple subtractions, the discussion is similar: the inclusion of the TPE will improve the reliability of the LHC, while the number of $N/D$-subtractions and of  DR contact-terms play analogous roles. 

 The study of situations where the long-distance potential is singular is in progress and might require further renormalization conditions.   This is also related with the previous discussion, since  TPE, as well as OPE in some partial-wave amplitudes,  lead to singular long-distance potentials, diverging faster than $1/r^2$ for $r\to 0$. The exact $N/D$ method is known to provide new solutions with respect to  the LSE scheme with a cutoff that is sent to infinity ($\Lambda\to\infty$), both for  attractive \cite{Entem:2016ipb} and repulsive \cite{Entem:2021kvs} singular potentials. In the future, we intend to study whether these additional $N/D$ solutions, which allow to reproduce higher parameters of the ERE,  can also be found within the LSE DR renormalization schemes, as done in this work for the $^1S_0$ regular OPE potential. As noted above, the main problem, when the LSE is solved in  the $\Lambda\to \infty$ limit, is that only one renormalization condition (singular attractive case) or none (singular repulsive case) can be used in each uncoupled partial wave \cite{Case:1950an,Frank:1971xx,PavonValderrama:2005gu,PavonValderrama:2005wv,PavonValderrama:2005uj}. A similar situation also appears in some scenarios of coupled partial wave amplitudes~\cite{PavonValderrama:2005wv,PavonValderrama:2005uj}.
This is a too restrictive renormalization scheme both for an accurate reproduction of the experimental data and for increasing the number of counterterms needed when the potential is calculated following a (chiral) power counting.

An open question is the design of a consistent power-counting relating the chiral order employed in the long-distance potential and the number of derivatives considered in the short-distance part of the interaction, beyond those strictly necessary to render UV finite the long-distance part of the amplitude. In the case studied in this work, the OPE can be renormalized with only one counter-term. Consideration of higher-order terms in the short-range potential has allowed us to fix other ERE parameters in addition to the scattering length. 

\acknowledgments
We thank E. Hern\'andez for his comments and clarifications. This work is part of the Grants PID2020-112777GB-I00, PID2022-141910NB-I00, PID2023-147458NB-C21,  PID2022-136510NB-C32 and CEX2023-001292-S of MCIN/AEI/10.13039/501100011033/, FEDER UE and MICIU/AEI/10.13039/501100011033 and of the Grants  CIPROM 2023/59 of Generalitat Valenciana  and  Junta de Castilla y Leon under Grant No. SA091P24.

\appendix
\section{Some results with a sharp cutoff and the relationship between the LSE DR scheme  and Ref.~\cite{Kaplan:1996xu} }

\subsection{Results with a sharp cutoff}
\label{app:cut}
In this Appendix, we first use a sharp cutoff $\Lambda$ to regularize the LSE amplitude of Eq.~\eqref{eq:tsol}. As discussed in the main text, the loop integral at threshold $J_0^R$ diverges logarithmically. It can be regularized by a cutoff $\Lambda$, 
\begin{figure}
\begin{center} 
\includegraphics[width=7.5cm]{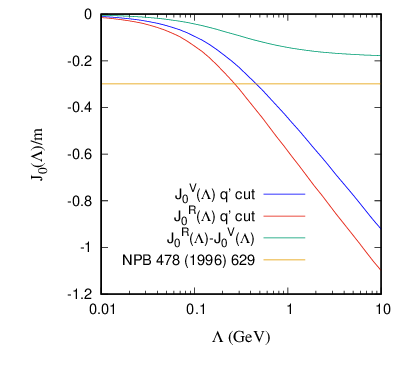}\includegraphics[width=7.5cm]{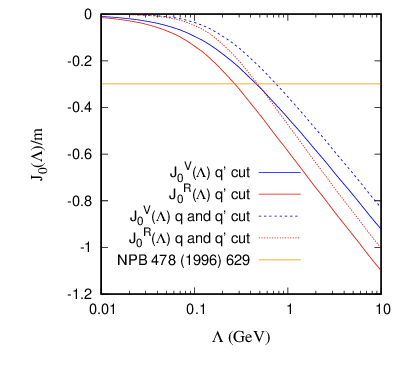}\\
\includegraphics[width=7.5cm]{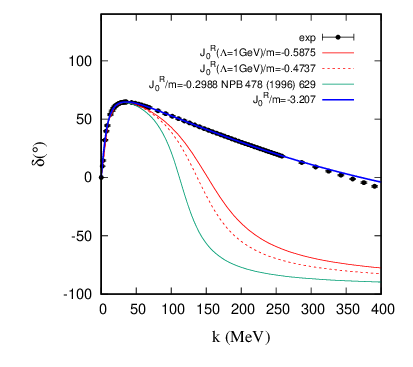}
\caption{\label{fig:J0div} Top: Different calculations of $J_0^R(\Lambda)$ using Eqs.~\eqref{eq:J0RTdiv} (solid red lines), \eqref{eq:J0RTdiv2} (dotted red line) and their respective leading approximations $J_0^V(\Lambda)$ (solid and dotted blue lines)  only accounting for the potential $V_\pi(q',q)$, as a function of the cutoff. The green curve in the left panel represents the difference $\left[J_0^R(\Lambda)-J_0^V(\Lambda)\right]$, when only one of the momentum integrations is truncated in the calculation, as in Eq.~\eqref{eq:J0RTdiv}. We also show the  $J_0^R/m =-0.2988$ horizontal line, which corresponds to the value that can be derived from the ${\overline{\rm MS}}$ renormalized coordinate-space propagator at threshold given in Fig.~A3 of Ref.~\cite{Kaplan:1996xu}, and that is obtained for $\Lambda\sim 270$ MeV [$\Lambda\sim 470$ MeV] using Eq.~\eqref{eq:J0RTdiv} [Eq.~\eqref{eq:J0RTdiv2}]. Bottom: $^1S_0$ $NN$ phase-shift calculations at ${\cal O}(Q^2)$ ($n=1$) for four different values of $J_0^R$. The solid blue and   red [dashed red] curves correspond to the original results of Ref.~\cite{Nieves:2003uu} and to the predictions with $J_0^R$  estimated using Eq.~\eqref{eq:J0RTdiv} [Eq.~\eqref{eq:J0RTdiv2}] and $\Lambda=1$ GeV, respectively. Finally, the green curve stands for the phase-shifts obtained using $J_0^R/m =-0.2988$, as inferred from \cite{Kaplan:1996xu}. In the last three cases the ERE parameters $a$ and $r$ are set to the values obtained in Ref.~\cite{Nieves:2003uu}, using Eq.~\eqref{eq:renororden2} with $\delta_{1,2}$ from Eq.~\eqref{eq:delta13org} to fix $g_{00}$ and $g_{01}$, while in the calculation of Ref.~\cite{Nieves:2003uu}, $J_0^R$ is also fitted to data below $k\leq 260$ MeV, leading to a more realistic $v_2$. The black dots with error bars are the experimental phase-shifts from Ref.~\cite{Stoks:1993tb}. }
\end{center}
\end{figure}
\begin{equation}
J_0^R\sim J_0^R(\Lambda) =   \int_0^\Lambda dq' \int_0^\infty dq\, T_\pi(q',q; k^2=0)\,.  \label{eq:J0RTdiv}
\end{equation}
Note that only one integral has been truncated since only when the second momentum is also integrated does the logarithmic divergence appear. The UV ill-defined contribution to $J_0^R$ arises from the double integration of the potential, where one can easily see that the first momentum integral is finite and the divergence appears in the subsequent integration~\cite{Nieves:2003uu}, 
\begin{eqnarray}
    \int_0^\Lambda dq' \int_0^\infty dq V_\pi(q',q) &=& -M_N\alpha_\pi \int_0^\Lambda \frac{dq'}{q'}\,\arctan\left(\frac{q'}{m_\pi}\right)= \nonumber-M_N\alpha_\pi 
    \left[\frac{\pi}{2} \ln \frac{\Lambda}{m_\pi}+{\rm Im}\, {\rm Li}_2\left(i \frac{m_\pi}{\Lambda} \right)\right]\\
    &=&   -M_N\alpha_\pi \left[\frac{\pi}{2} \ln \frac{\Lambda}{m_\pi}+\frac{m_\pi}{\Lambda} + {\cal O}\left(\frac{m_\pi^3}{\Lambda^3}\right)\right] \label{eq:J0Rdiv}
    \end{eqnarray}
where ${\rm Li}_2(z)=\sum_{s=1}^\infty z^s/s^2$ is the di-logarithm function. The bare couplings $g_{ms}$ in Eq.~\eqref{eq:vs} will depend on the cutoff $\Lambda$ and should be tuned to reproduce the cutoff-independent physical $\delta_i$ LECs introduced in Eq.~\eqref{eq:deltas}.

Instead, one might cut both momentum integrations and evaluate $J_0^R$ as
\begin{equation}
J_0^R\sim J_0^R(\Lambda) =   \int_0^\Lambda dq' \int_0^\Lambda dq\, T_\pi(q',q; k^2=0)\,.  \label{eq:J0RTdiv2}
\end{equation}
Estimating $J_0^R$ using  Eq.~\eqref{eq:J0RTdiv} or  Eq.~\eqref{eq:J0RTdiv2}, and computing the finite part of the loop function $\bar J(k)$ completely free of finite cutoff effects, is in the spirit of the renormalization procedure adopted in Ref.~\cite{Nieves:2024dcz}.  However, the DR LSE amplitude of Eq.~\eqref{eq:tsol} with $J_0^R$ calculated using Eq.~\eqref{eq:J0RTdiv} or Eq.~\eqref{eq:J0RTdiv2} does not correspond to the LSE solution with a sharp cutoff. The latter would amount to taking $V(p',p)\Theta(\Lambda-p')\Theta(\Lambda-p)$ in the LSE of Eq.~\eqref{eq:def-lse}, with $\Theta$ the Heaviside function, which would, for instance, induce unnecessary/inappropriate cutoff dependencies on the finite $T_\pi(q',q)$ long-distance amplitude. It would neither correspond to regularizing only the short-distance potential, i.e. $V(p',p)\to V_\pi(p',p)+V_s(p',p)\Theta(\Lambda-p')\Theta(\Lambda-p)$. In that case, the cutoff would not affect $T_\pi(q',q)$, but it would appear in other parts of the calculation such as ${\rm Re}\bar J_0(k)$ or $L_0(k)$, or worse,  it would make the relations in Eq.~\eqref{eq:vpi-rel} no longer valid, and therefore Eq.~\eqref{eq:tsol} would not be formally obtained.

In the upper panels of Fig.~\ref{fig:J0div}, we show different calculations of $J_0^R(\Lambda)$ using Eqs.~\eqref{eq:J0RTdiv}, \eqref{eq:J0RTdiv2} and their respective leading approximations only accounting for the potential $V_\pi(q',q)$ as in Eq.~\eqref{eq:J0Rdiv}. The latter [$J_0^V(\Lambda)$] contain the UV divergence of $J_0^R(\Lambda)$, and the differences between the two are finite and independent of the cutoff as it grows (see, for example, the green curve in the top left panel). On the other hand, the estimates for $J_0^R$ of Eq.~\eqref{eq:J0RTdiv} and Eq.~\eqref{eq:J0RTdiv2} are significantly different for small and moderate values of $\Lambda$, and only agree within 10\% for sufficiently large cutoffs, at least above 10 GeV. 

We notice that $J_0^R$ understood as an indeterminate subtraction constant, as in the main text, can take any real value. However, some value regions, for example $J_0^R>0$,  might not be reached when approximating $J_0^R$ using the threshold loop function computed with a sharp cutoff. 

In the bottom panel of Fig.~\ref{fig:J0div}, we show four $^1S_0$ $NN$ phase-shift calculations at ${\cal O}(Q^2)$ ($n=1$) for different values of $J_0^R$. Three of them correspond to  sharp-cutoffs in the range of 300 MeV to 1 GeV, which would be natural in the so-called chiral unitarity approaches~\cite{Oller:1997ti, Oset:1997it, Oller:2000fj, Kaplan:1996xu}. We observe that they can only reproduce the Nijmegen phase-shifts below $k=50-75$ MeV, and fail to describe the slow data decrease for higher momenta. This is in sharp contrast with the phase-shift calculation of Ref.~\cite{Nieves:2003uu},  where $J_0^R$ was fitted and provided a surprising good description of the data up to momenta around $300-350$ MeV.

As noted in the main text, a nonzero value for $J_0^R$ induces a correlation between the higher order terms of the ERE expansion that is governed by the inverse of this finite $J_0^R$ LEC rather than by the dispersion length, which would not be appropriate in cases where the latter is abnormally large, as in the $^1S_0$ $NN$ sector~\cite{Kaplan:1996xu}. The numerical value for $J_0^R$ obtained in the ${\cal O}(Q^2)$ calculation of Ref.~\cite{Nieves:2003uu} was around $-3.2 m$ and cannot be associated with any meaningful cutoff. Actually, as we will see in the next subsection, the well-known KSW approach of Ref.~\cite{Kaplan:1996xu} proposed to modify the expansion and rely on the  inverse amplitude method (IAM) \cite{Dobado:1989qm,GomezNicola:2007qj} to describe the data, which is an indication that there exist physics beyond the unitarity logarithms playing a relevant role. This might have some similarities with the case of the $\rho-$resonance in $\pi\pi$ scattering, where large ${\cal O}(p^4)$ Gasser-Leutwyler LEC contributions show up that cannot be approximated by computing the loop function with a natural-size cutoff, in contrast to what occurs in the scalar-isoscalar $f_0(500)$ sector~\cite{Oller:1997ti, Nieves:1998hp,Nieves:1999bx,Nieves:2009ez, Nieves:2011gb}. The value of $J_0^R$ in \cite{Nieves:2003uu} is determined mainly by the experimental $v_2$ ERE parameter, and  forces that $T(k)=T_\pi(k)$ for $k=\sqrt{\delta_1/\delta_2}\sim 310$ MeV (see Eq.~\eqref{eq:serie1} and Fig.~\ref{fig:forder1}) and allows the predicted phase-shift to become negative beyond 300 MeV, in agreement with the Nijmegen data. In fact, the results in the lower panel of Fig.~\ref{fig:J0div} seem to indicate that the large value of $|J_0^R/m|$ is related to the slow decrease of the phase shift and the position of the zero of the amplitude ($\delta=0$), and not to the existence of the virtual state.

\subsection {${\cal O}(Q^2)$ DR scheme and Ref.~\cite{Kaplan:1996xu}}
\label{app:ksw}
We point out here the similarity between some of the results obtained by Kaplan et al. in Ref.~\cite{Kaplan:1996xu}, by solving the Schr\"odinger equation in coordinate space, and the ${\cal O}(Q^2)$ DR amplitude found in Ref.~\cite{Nieves:2003uu} and used in this work. Actually Eq.~(4.16) in Ref.~\cite{Kaplan:1996xu} totally agrees with Eq.~\eqref{eq:tsol} at ${\cal O}(Q^2)$, taking into account the normalization change between the amplitudes used here $T$ and in that reference ${\cal A}=-\pi^2 T/m$. In addition, the square of the OPE wave function at the origin  $[\chi_{\bf k}( \bf 0)]^2$, the renormalized coordinate-space propagator from the origin to the origin in the presence of the OPE $\tilde G_E({\bf 0},{\bf 0})$, the contact couplings $\tilde C$ and  $\tilde C/(2\Lambda^2)$ in \cite{Kaplan:1996xu} are  here $[1+L_0(k)]^2$, $m[-i\pi k/2+J_0^R+\bar J_0(k)]/\pi^2$, $\pi^2 g_{00}/m$ and  $\pi^2 g_{01}/m$, respectively. Actually, we have reproduced Figs. A2 and A3 from~\cite{Kaplan:1996xu}, taking into account in the latter that the value of the ${\overline{\rm MS}}$ renormalized coordinate-space propagator at threshold given there determines $J_0^R/m \sim -0.2988$ (displayed in the upper plots of Fig.~\ref{fig:J0div}). Finally, we have also reproduced the dashed-dotted curve of Fig.~7 using Eq.~\eqref{eq:renororden2} and $\delta_{1,2}$ from Eq.~\eqref{eq:delta13org} to fix $g_{00}\sim -0.4263/m$ and $g_{01}\sim 27.3845/m^3$, which lead to values for  $C_{\overline{\rm MS}}(\mu=m_\pi)$ and $\Lambda^2_{\overline{\rm MS}}(\mu=m_\pi)$ in good agreement with those quoted in Eq.~(4.17) of Ref.~\cite{Kaplan:1996xu}\footnote{For simplicity, we have taken for the  $\delta_0$ and $\delta_1$ LECs the values given in Eq.~\eqref{eq:delta13org}, since they lead to $a$ and $r$ ERE parameters  very close to those used in \cite{Kaplan:1996xu}, deduced from the Nijmegen data.}. These phase-shifts are shown by the green curve in the bottom panel of Fig.~\ref{fig:J0div}.

We note that Eq.~\eqref{eq:tsol} derived here extends the ${\cal O}(Q^2)$ formulae of Refs~\cite{Nieves:2003uu, Kaplan:1996xu}, since  Eq.~\eqref{eq:tsol} includes ${\cal O}(Q^6)$ contact terms, and its extension to higher orders is straightforward. 

A new power counting scheme is also proposed in Ref.~\cite{Kaplan:1996xu} leading to their Eq.~(4.20),  which provides a substantial better description of the  $^1S_0$ $NN$ Nijmegen phase shits~\cite{Stoks:1993tb} than Eq.~(4.16) [see the solid curve of Fig. 7 of Ref.~\cite{Kaplan:1996xu}]. This new power counting is just the result of applying the IAM to the momentum expansion of the contact amplitude, including the finite OPE in the LO term. This is so because  Eq.~(4.20) of Ref.~\cite{Kaplan:1996xu}  amounts to
\begin{equation}
 T_{\rm IAM}(k) = \frac{\left[T_0(k)\right]^2}{T_0(k)-T_1(k)} \label{eq:TIAM}
\end{equation}
where $T_0(k)=T_{{\cal O}(Q^0)}(k) = T_\pi(k)-\left[1+L_0(k)\right]^2/\left(g^{-1}_{00}+i\pi k/2 -\bar J_0(k) -J_0^R\right)$ is the zero-order amplitude studied in Subsec.~\ref{order0}, and the sub-leading contribution $T_1(k)$ reads
\begin{equation}
 T_1(k) = \frac{2g_{01} (k_\pi^2-k^2)\left[1+L_0(k)\right]^2}{g_{00}^2\left(g^{-1}_{00}+i\pi k/2 -\bar J_0(k) -J_0^R\right)^2}
\end{equation}
Note that in this order $T_{\rm IAM}(k)$ in Eq.~\eqref{eq:TIAM} depends only on the combinations $(g^{-1}_{00}-J_0^R)$ and $g_{01}/g^2_{00}$ and not separately on $g_{00},g_{01}$ and $J_0^R$. However, the numerical values of the renormalized couplings $g_{00}$ and $g_{01}$, and hence of $C_{\overline{\rm MS}}(\mu=m_\pi)$ and $\Lambda^2_{\overline{\rm MS}}(\mu=m_\pi)$, will be sensitive to the choice of the threshold loop function $J^0_R$, or equivalently of the threshold $\overline{\rm MS}$ Green function at the origin $\tilde G^{\overline{\rm MS}}_{E=0}({\bf 0},{\bf 0})$. 

We observe that $T_{\rm IAM}(k) $ satisfies exact unitarity due to the fact that $T_1(k)/\left[T_0(k)\right]^2$ is real for $k>0$, as can be easily shown using Eqs.~\eqref{eq:l0} and \eqref{eq:imJ0}.   On the other hand, $T_{\rm IAM}(k)$ does not have the correct OPE LHC, unlike the ${\cal O}(Q^2)$-DR amplitude which inherits the exact one from $T_\pi$. As a further consequence, $T_{\rm IAM}(k)$ does not match any $N/D_{nd}$ solution. 

We end by reiterating that in this work, we do not study the interesting question of power-counting. The main goal of our work is to show that the solution of the LSE amplitude in Eq.~\eqref{eq:tsol} reproduces the exact  $N/D_{nd}$ calculation, with the appropriate number and typology of subtractions to obtain the effective range parameters taken as input to renormalize the LSE amplitude.

\bibliographystyle{apsrev4-1}
\bibliography{dimreg}

\end{document}